\documentclass[usenatbib,twocolumn]{mnras}
\usepackage{graphicx}
\usepackage{color}
\usepackage{setspace}
\usepackage{amsmath,amssymb,amstext}
\usepackage{dcolumn}
\newcolumntype{d}[1]{D{.}{.}{#1}}
\usepackage[multiple]{footmisc}

\makeatletter
\newlength{\abovecaptionskip}%
\makeatother
\usepackage{threeparttable}
\usepackage{multicol}
\usepackage{pdflscape}

\def\apj{\rm ApJ}
\def\apjl{\rm ApJL}
\def\apjs{\rm ApJS}
\def\aj{\rm AJ}
\def\jcap{\rm JCAP}
\def\mnras{\rm MNRAS}
\def\nat{\rm Nature}

\def\pasp{\rm PASP}

\def\aap{\rm AAP}
\def\araa{\rm ARA\&A}

\def\rmxaa{\rm RMXAA}

\def\lya{Ly$\alpha$}
\def\hcmpc{$h^{-1} \mbox{ cMpc}$}

\defcitealias{Hennawi2020}{\scshape H2020}

\begin{document}

\title [IGM metallicity and the C\,IV forest]{Constraining IGM enrichment and metallicity with the \ion{C}{\uppercase{IV}} forest correlation function}
\author[Tie et al. ]{Suk Sien Tie$^1$,
    Joseph F. Hennawi$^{1,2}$, 
    Koki Kakiichi$^1$
    and Sarah E. I. Bosman$^3$,
    \\
  $^{1}$ Department of Physics, Broida Hall, University of California, Santa Barbara, Santa Barbara, CA 93106-9530, USA \\
  $^{2}$ Leiden Observatory, Leiden University, Niels Bohrweg 2, 2333 CA Leiden, Netherlands \\
  $^{3}$ Max-Planck-Institut f\"{u}r Astronomie, K\"{o}nigstuhl 17, D-69117 Heidelberg, Germany
   }

\maketitle

%% -- An intro sentence on why studying cosmic metals is important, and important for the EOR. 
%% -- At high-z metallicity measurements are hardly available. At even higher redshifts limited sensitivity means
%% you cannot probe metals in densities characteristic of the IGM. 
%% -- We present a new method that gets around the point above by statistically averaging over all the pixels to detect
%% extremely weak metal line absorption that would otherwise not be accessible. 

\begin{abstract}
The production and distribution of metals in the diffuse intergalactic medium (IGM) have implications for galaxy formation models, cosmic star formation history, and the baryon (re)cycling process. 
Furthermore, the relative abundance of metals in high versus low-ionization states has been argued to be sensitive to the Universe's reionization history. However, measurements of the background metallicity of the IGM at $z\sim 4$ are sparse and in poor agreement with one another, and reduced sensitivity in the near-IR implies that
probing IGM metals at $z > 4$ is currently out of reach if one adheres to the traditional method of detecting individual absorbers. We present a new technique based on clustering analysis that enables the detection of these weak IGM absorbers by statistically averaging over all spectral pixels, here applied to the \ion{C}{IV} forest. We simulate the $z=4.5$ IGM with different models of inhomogeneous metal distributions whereby halos above a minimum mass enrich their environments with a constant metallicity out to a maximum radius. We generate mock skewers of the \ion{C}{IV} forest and investigate its two-point correlation function (2PCF) as a probe of IGM metallicity and enrichment topology. The 2PCF of the \ion{C}{IV} forest demonstrates a clear peak at a characteristic separation corresponding to the doublet separation of the \ion{C}{IV}  line. The peak amplitude scales quadratically with metallicity, while enrichment morphology affects both the shape and amplitude of the 2PCF. The effect of enrichment topology can also be framed in terms of the metal mass- and volume-filling factors, and we show their trends as a function of the enrichment topology. For models consistent with the distribution of metals at $z \sim 3$, we find that we can constrain [C/H] to within 0.2 dex, log$\,M_{\rm{min}}$ to within 0.4 dex, and $R$ to within 15\%. While the correlation function can be overwhelmed by the strongest absorbers arising from the circumgalactic medium of galaxies, we show how that these strong absorbers can be easily identified and masked, allowing one to recover the underlying IGM signal. The auto-correlation of the metal-line forest presents a new and compelling avenue to simultaneously constrain IGM metallicity and enrichment topology with high precision at $z > 4$, thereby pushing such measurements into the Epoch of Reionization. 

%% SST omitted
%% JFH2 This statement is vague about different methods, I would consider omitting it. 
%When viewed alongside measurements obtained using different methods, this will help us converge towards a consistent picture of the Universe's chemical enrichment history. 
\end{abstract}


\section{Introduction}
The existence of heavy elements, or metals, in the intergalactic medium (IGM) has been known for decades via absorption line studies of background quasars. Beginning with observations of \ion{C}{IV} and \ion{S}{IV} at $z\sim3$ (e.g. \citealp{Cowie1995, SongailaCowie1996, Songaila2001, Ellison2000,Schaye2007}), other metals such as \ion{O}{VI} and Mg II have also been observed (e.g. \citealp{Schaye2000,Carswell2002,Simcoe2002,Bergeron2002,Pieri2010,Chen2017}). At $z < 1$, \cite{Cooksey2010} 
%% SST done
%% JFH I would probably not call z < 1 the local neighborhood. Perhaps just say at z < 1 or at low-z or something like that. Local
%% neighborhood usually applies to nearby galaxies, etc. 
have detected \ion{C}{IV} in the spectra of quasars in the UV, while myriad observations in the near-infrared have detected \ion{C}{IV} up to $z\sim6$ (e.g. \citealp{Songaila2005,Becker2009,Ryan-Weber2006,Ryan-Weber2009,Simcoe2006,Simcoe2011b,Codoreanu2018}). In parallel, numerical simulations indicate that the $z \sim 2-3$ IGM has a typical metallicity of [C/H] $\sim$ $-3$ to $-2$ \citep{Haehnelt1996,Rauch1997,Dave1998,Carswell2002,Bergeron2002}.

The production, transport, and distribution of metals in the IGM are inextricably tied to galaxy formation; while fuel from
%% SST done
%% JFH while fuel from the IGM seeds ...
the IGM seeds the birth and growth of galaxies, feedback processes in galaxies are believed to recycle materials back into the IGM. Models can be constructed where the dominant enrichment mechanism is galactic winds and/or Population III stars, with different implications for the metal distribution (e.g. \citealp{CenOstriker1999,Aguirre2001a,Aguirre2001b,Madau2001,Theuns2002,Scannapieco2003,Aguirre2005,OppenheimerDave2006,Pieri2006,Kobayashi2007,CenChisari2011}). Observations of metals in the IGM therefore constrain the nature of this enrichment process and models of galaxy formation.

Metal absorption lines at high redshifts are probes of reionization. By virtue of their low oscillator strengths and similar ionization energy to \ion{H}{I}, low-ionization metals lines such as \ion{O}{I} and \ion{Si}{II} are good tracers of neutral hydrogen in the pre-reionized IGM \citep{Oh2002}, allowing one to constrain the reionization topology and history. As reionization progresses, overdense regions that are predominantly neutral should produce forests of these low-ionization lines, which should gradually disappear and make way for forests of high-ionization lines like \ion{C}{IV} and \ion{O}{VI} at the end of reionization, due to the increasingly hard UV background. The transition from forests of high-ionization to low-ionization metal absorption lines as redshift increases would be the hallmark of reionization, analogous to the emergence of the Gunn-Peterson trough in the Ly$\alpha$ forest of hydrogen. By simulating the \ion{Mg}{II} forest at $z$=7.5, \cite{Hennawi2020} shows that the \ion{Mg}{II} metallicity constrains the hydrogen neutral fraction. 

%% SST included more details and redshifts for measurements.
%% JFH This paragraph would be better written if it included more numbers and quantitative information. For exampnle, what
%% signal to noise and resolution are you referring too? What column densities of e.g. CIV, and putting in the actual numbers from 
%% abstracts for e.g. the b/g IGM metallicity measurements. I think you also need to quote the redshifts of these measurements and highlight that the
%% the highest redshift at which metallicity can be measured is z = 4.X from Simcoe et al. 2011

%% SST not sure where to best place this sentence, but I've moved it to end of P1. 
%% JFH2 This title sentence seems inappropriate in the paragraph below. The paragraph that %% is about measurements of IGM meatallicity at various redsfhits, so why start
%% the paragraph with a title sentence about simulations, since this is the only mention
%% of simulations. Maybe move this up to the previous paragraph. 
%Numerical simulations indicate that the $z \sim 2-3$ IGM has a typical metallicity of [C/H] $\sim$ $-3$ to $-2$ \citep{Haehnelt1996,Rauch1997,Dave1998,Carswell2002,Bergeron2002}. 
Despite quasar observations spanning wide redshift ranges, measurements of the IGM metallicity are mostly concentrated at $z\sim3$, where high resolution (FWHM $<$ 20 km/s) and high SNR ($>20$) measurements are most easily attainable with current ground-based telescopes, such as with Keck's High Resolution Echelle Spectrometer (HIRES; \citealp{Vogt1994}), VLT's ESO UV-visual echelle spectrograph (UVES; \citealp{Dekker2000}), and Magellan's Folded port InfraRed Echellette (FIRE; \citealp{Simcoe2013}). High sensitivity measurements are needed to detect and resolve the weak metal lines in the low-density IGM, otherwise observations will instead be dominated by high column density ($N_\mathrm{\ion{H}{I}} > 10^{14.5}$ cm$^{-2}$) absorbers in the circumgalactic medium (CGM). Using standard Voigt profile fitting, the carbon abundance in the IGM has been measured to be [C/H] $\sim -2.5$ at $z=3-3.5$ for absorbers with $N_\mathrm{\ion{H}{I}} \gtrsim 10^{12}$ cm$^{-2}$  \citep{Songaila1997,Ellison2000} to [C/H] = $-3.55$ at $z=4.25$ for absorbers with $N_\mathrm{\ion{H}{I}} \geq 10^{14.5}$ cm$^{-2}$ \citep{Simceo2011a}, which is the highest redshift measurement of the IGM metallicity. When compared with lower-redshift \ion{O}{VI} and \ion{C}{IV} measurements \citep{Simcoe2004}, \cite{Simceo2011a} further concluded that the carbon abundance decreased moderately towards higher redshift, suggesting that almost half of the metals were deposited between $z=4$ and $z=2$. Using a more statistical appproach known as the pixel optical method that measures
%% SST change to measures
%% JFH2 tabulats is an odd word choice here. Maybe "estimates?"
the distribution of \ion{C}{IV} optical depth as a function of the H I optical depth in the same gas \citep{CowieSongaila1998,Aguirre2002,Aguirre2004,Aguirre2008}, \cite{Schaye2003} measured a similar median [C/H] = $-3.47$ at $z=3$ but found very little to no evolutionary trend between $z=4.1$ to $z=1.8$ \citep[see also][]{Aracil2004}.
%% SST done
%% JFH2 The Aracil citation formatting is incorrect. It should be \citep[see %% also][]{Aracil2004}. 
This instead suggests that majority of the metals were deposited predominantly
%% SST done
%% JFH I'm not sure singularly is the right adjective here, perhaps predominantly? Maybe highlight the apparent contradiction with the
%% the evolution of the cosmic SFR evolution?
at high redshifts, in contrast to the evolution of the cosmic star formation history.
%% SST done, added a few more relevant ones
%% JFH Can you double check that you cited all the papers by e.g. Schaye and Aguiree on this topic. Since Joop is now my colleague
%% it is a bit embarassing if I fail to cite his past work, and he did a lot of work on this subject. Thanks!

%% JFH IMPORTANT: You are missing an paragraph in the intro that motivates this new approach. This is rather important and without it your paper does not
%% have a strong motivation. I think the motivations are 1) The traditional Voigt profile approach breaks down when the Ly-a forest can no longer be easily
%% decomposed into Voight profiles, i.e. line blanketing, 2) at high-z the sensitivity and resolution are usually worse becuase the relevant transitions 
%% are shifting into the IR. As a result individual CIV absorbers arising from around the mean IGM density are exceedingly dificult to detect. Our statitsical
%% approach of correlation funciton measurement gets around this limitation. 3) Are there other motivations? 
%% JFH To make this motivation more compelling you probably
%% need to beef up the (somewhat weak) paragraph above on past metallicity measurements. You probably also need to make abundantly clear that nearly all z > 4 
%% CIV detections/studies are CGM and not IGM absorbers, given their column densities. 

Existing measurement methods outlined above become impractical at higher redshifts, as they all rely on detecting \lya\ lines. At $z > 4$, the combination of large scattering cross-section and residual neutral fractions in the IGM cause \lya\ absorption lines to become
%% JFH2 Actually a subset of the lines are saturated at all redshifts so saturation alone
%% is not per se the issue, although part of it. The real issue is line blanketing
%% because a large fraction of the volume contains gas with optical depth large
%% enough to be saturated. 
saturated and line-blanketed, making it impossible to decompose them into individual lines and to obtain accurate column density measurements. It is also challenging to detect metal lines at higher redshifts as they shift closer into the NIR (e.g. \ion{C}{IV} redshifts to 8520 \r{A} at $z=4.5$), where the sky background is higher, and at $>$ 1 $\mu$m, the detector sensitivity and resolution are worse. 
%% SST changed
%% JFH2 Telescopes are just as sensitive. The issue is the sky background which is higher. %% It is only approaching 1micron where detector sensitivity starts to decrease.
%% Resolution is only worse if you work > 1um in the near-IR. This is mostly because 
%% higher sky background means you don't have enough S/N to work at high resolution.
Given the strict observational requirements to detect IGM absorbers ($N_\mathrm{\ion{H}{I}} \lesssim 10^{13}$ cm$^{-2}$) at high redshifts and the limitations of current methods, most detections of metal absorbers likely originate from overdense gas in the CGM. To push measurements to higher redshifts, a new statistical method that does not require anchoring on \lya\ lines is much needed. 

%% JFH I'd suggest moving this pargraph down one, and moving the one on triply-ionized Carbon up.
We present a new technique that analyses the clustering of the metal line forest to probe the IGM metallicity and enrichment morphology. This is an extension of the \cite{Hennawi2020} (hereafter \citetalias{Hennawi2020}) 
%% SST done
%% JFH Maybe clarify that this paper was focused on MgII
method that focuses on the \ion{Mg}{II} forest at $z = 7.5$, but we applied it here to the \ion{C}{IV} forest at $z = 4.5$. We present results for patchy metal distributions as opposed to a uniform metal distribution in the original work. 
Previous work studying the clustering of metal absorbers, especially \ion{C}{IV} absorbers, focus on the clustering of discrete absorption systems. These studies establish that \ion{C}{IV} absorbers cluster strongly at velocity separations $\Delta v \lesssim$ 500 km/s, while being uncorrelated
%% SST changed to uncorrected
%% JFH2 clarify what is flat. I think you mean the correlation function is flat, but this
%% is ambiguously constructed. 
on larger scales \citep{Sargent1980,Steidei1990,PetitjeanBergeron1994,Rauch1996,Pichon2003}. A more comprehensive study by \cite{Boksenberg2015} (which is an updated work of \citealp{Boksenberg2003}), comprising 1099 \ion{C}{IV} absorber components in 201 systems spanning 1.6 $\lesssim z \lesssim$ 4.4, shows that \ion{C}{IV} components exhibit strong clustering on scales $\Delta v < $ 300 km/s, with most clustering occuring at $\Delta v < $ 150 km/s. They conclude that the detected clustering is a result of the peculiar velocities of gas clouds and the clustering of components within each system (i.e. cloud-cloud clustering in the CGM of galaxies), as opposed to a result of galaxy clustering, where clustering on larger scales is expected. 
%% SST omitted
%% JFH2 I don't think this sentence on MgII absorbers is necessary. We are talking CIV
%% here. 
%Studies of \ion{Mg}{II} absorbers find similar small-scale clustering \citep{PetitjeanBergeron1990,Churchill2003}. 
The work that is closest in motivation and method to ours is \cite{Scannapieco2006}, 
%% SST done
%% JFH2 Can you also add a (see also Martin et al. 2010), I'm on that paper
%% https://ui.adsabs.harvard.edu/abs/2010ApJ...721..174M/abstract
who measured the clustering of \ion{C}{IV}, \ion{Si}{IV}, \ion{Mg}{II}, and \ion{Fe}{II} absorbing components over $z = 1.5 - 3$ in 19 quasar spectra to constrain the IGM metallicty and enrichment topology (see also \citealp{Martin2010} for a similar method applied to binary quasar spectra). They found that the \ion{C}{IV} correlation function is inconsistent with a model where the IGM metallicity is constant or a power-law function of overdensity and more in line with a model where metals are trapped within bubbles of radius $\approx\,$2 cMpc around $\approx\,$10$^{12} M_{\odot}$ halos at $z = 3$. However, their conclusion disagrees
%% SST changed
%% JFH2 direct contrast is an awkward construction. direct conflict? in disagreement with?
with that of \cite{Booth2012}, who found that the $z = 3$ IGM is predominantly enriched by low-mass ($< 10^{10} M_{\odot}$) galaxies out to $r \geq 100 $ proper kpc, by comparing simulations with the pixel optical depth measurements of \cite{Schaye2003}.

In contrast to previous work, we focus on the clustering of transmitted flux in the \ion{C}{IV} forest, treating the flux field as a continuous random field as opposed to
a collection of discrete  absorbers. This measurement paradigm, 
which does not require identifying individual lines/absorbers, is common in studies of the baryon acoustic oscillations (BAO) using the \lya\ forest (e.g., \citealp{Bautista2017,Bourboux2017,Bourboux2020}). The \lya\ forest is a tracer of choice at $z > 2$ but it leaves the optical window at lower redshifts. At $z < 2$, the metal-line forests are viable probes of large-scale structures; both the \ion{C}{IV} and \ion{Mg}{II} forests have been cross-correlated with low-redshift quasars and galaxies using BOSS/eBOSS data to constrain BAO parameters \citep{Pieri2014,Zhu2014,Perez2015,Blomqvist2018,Gontcho2018,Bourboux2019}.

%% JFH for other relevant contaminanting lines you can see the list here: https://github.com/profxj/xidl/blob/master/Spec/Lines/Lists/lls.lst
%% Note that the third column is oscillator strength, although the right way to do this would be to weight oscillator strength 
%% by solar abundance, but actually you care most about line density (which also depends on e.g. ionization state) 
There are several reasons why triply-ionzed carbon (\ion{C}{IV}) is used as the tracer of choice for our work. First and foremost, carbon is one of the most abundant metal elements in the Universe, after oxygen
($N_{\rm C}/N_{\rm H}$ = 2.95 $\times$ 10$^{-4}$ and $N_{\rm O}/N_{\rm H}$ = 5.37 $\times$ 10$^{-4}$) \citep{Asplund2009}.
%% SST changed
%% JFH2 I would use the numbers relative to hydrogen as they are listed in Draine, i.e. 
% n_C/n_H. These metallicity units that you are using here which I always forget are 
%% rather confusing. You can see e.g. the  Table in the first chapter of Bruce Draine's ISM book. 
%% SST done
%% JFH It might be useful to quote the solar abundance here as a reference in terms of number. 
Most carbon in the IGM is expected to be in the triply-ionized state due to ionization by the UV background. The \ion{C}{IV} line is also the dominant absorption line on the red side of the \lya\ forest. Besides potential contamination from lower-redshift \ion{Fe}{II} $\lambda1608$\r{A}, \ion{Al}{III} $\lambda1854$\r{A} and $\lambda1862$\r{A}, and \ion{Mg}{II} $\lambda2796$\r{A} and $\lambda2804$\r{A} lines, 
%% SST done
%% You might also want to mention AlIII as a contaminant
it does not suffer from foreground contamination from other common metal lines, as opposed to bluer lines like \ion{Si}{IV} $\lambda1394$\r{A}.
%% SST removed NV
%% JFH There is a typo above. SiIV is is at 1393 and 1402, not NV. NV is so rare that I would not even bother to mention it. 
\ion{C}{IV} has a strong doublet feature at rest-frame wavelengths $\lambda1548.20$\r{A} and $\lambda1550.78$\r{A}, which will result in a strong correlation peak at the velocity separation of the doublet at 498 km/s, thus lending itself naturally to a
%% SST clarified above
%% JFH It may not be so obvious why this lends itself naturally, i.e. you might clarify 
correlation function analysis. These wavelengths redshift to $\sim$ 8520 \r{A} at $z=4.5$, conveniently between the onset of atmospheric telluric
%% SST done
%% JFH telluric
absorption bands. 

In \S \ref{methods} we describe the simulation used in this work and how we generate metal distributions and create \ion{C}{IV} skewers. We compute the correlation function of the \ion{C}{IV} forest in \S \ref{results} and describe how it varies with model parameters. We will show that the correlation function has a sensitive dependence on the IGM metallicity and enrichment topology. We also investigate the effects of 
%% SST done
%% JFH contamination by
contamination by CGM absorbers and present methods that can remove them effectively and so recover the underlying IGM signal. In \S \ref{inference} we estimate the expected precision of the inferred model parameters using mock observations and show that they can be constrained to relatively high precision. Finally, we discuss our results and conclude in \S \ref{conclusion}. 

Throughout this paper, our mock observations are made up of 20 quasar spectra assuming a \ion{C}{IV} forest pathlength of $dz=1.0$ per quasar, resulting in a total pathlength of $\Delta z$ = 20. The spectra are convolved with a Gaussian line spread function with FWHM = 10 km/s ($R$=30,000; achievable with Keck/HIRES or VLT/UVES), where our spectral sampling is 3 pixels per resolution element. Gaussian random noise with $\sigma = $ (SNR)$^{-1}$ and SNR = 50 are added to each pixel. 
%% SST added
%% JFH2 You need to state the number of pixels per resolution element to completely specify
%% the forward model. Might was sell do that here. 
Throughout this work, we adopt a $\Lambda$CDM cosmology with the following parameters: $\Omega_m$ = 0.3192, $\Omega_\Lambda$ = 0.6808, $\Omega_b$ = 0.04964, and $h$ = 0.67038, which agree with the cosmological constrains from the CMB \citep{Planck2020} within one sigma. All distances in this work are comoving, denoted as cMpc or ckpc, unless explicitly indicated otherwise. We define metallicity as $[\mathrm{C/H}]$ $\equiv \mathrm{log}_{10}(Z/Z_{\odot})$, where $Z$ is the ratio of the number of carbon atoms to the number of hydrogen atoms and $Z_{\odot}$ is the corresponding Solar ratio, where we used $\mathrm{log}_{10}(Z_{\odot}) = -3.57$ \citep{Asplund2009}.
%% SST done
%% JFH Some important references are missing, like this one: 
%% https://arxiv.org/pdf/1801.01852.pdf
%% Also I think you need to add a review of efforts to study the clustering of CIV absorbers. I'm pretty sure
%% I reviewed these studies in my MgII forest paper in the section on the CGM. Although different in detail, 
%% these papers are similar and spirit and need to be discussed. 

\section{Methods}
\label{methods}
\subsection{Simulation}
\label{simulation}
We use the $\texttt{Nyx}$ code \citep{Almgren2013,Lukic2015} to simulate the \ion{C}{IV} forest and analyse the output at $z=4.5$. The $\texttt{Nyx}$ code is an adaptive mesh-refinement (AMR) $N$-body + hydrodynamics code that is specifically designed to simulate the IGM, capturing gravitational interactions while allowing the direct modeling of heating and cooling processes in the IGM. It models radiative processes assuming an optically-thin medium and uses the UV background (UVB) prescription from \cite{HM2012}. Our simulation assumed $\Lambda$CDM cosmology with $\Omega_m$ = 0.3192, $\Omega_\Lambda$ = 0.6808, $\Omega_b$ = 0.04964, $h$ = 0.6704, $\sigma_8$ = 0.826 and $n_s$ = 0.9655, which agree with the cosmological constrains from the CMB \citep{Planck2020} within one sigma. Initial conditions were generated using the MUSIC code \citep{HahnAbel2011} with a transfer function generated by CAMB \citep{Lewis2000,Howlett2012}. The simulation is started at $z=159$ and instantaneously reionized at $z=6$ when a spatially uniform and time-varying UVB is abruptly turned on. Radiative feedback is computed via an input list of photoionization and photoheating rates that vary with redshift. Our simulation box has a total grid size of 4096$^3$ and a length of 100 \hcmpc\ (12934.987 km/s) on each side, which gives a grid scale of $24$ $h^{-1}$ 
%% SST done
%% JFH the \sim symbol means order of magnitude. Don't use it to represent =. If you know the number, don't use \sim
ckpc (3.16 km/s). The baryon density, temperature, and peculiar velocity are output at each grid cell.
%% SST done
%% JFH Clarify somewhere in the text whether you are quoting comoving or proper distances. 
 
\subsection{Spatial distributions of metals}
\label{spatmetal}
Besides galaxy formation simulations 
%% SST done
%% JFH "full simulations" is ambiguous and vague. I would say simulations that model galaxy formation 
that model the creation and transport of metals self-consistently, common ways to generate the spatial distributions of metals in post-processing include creating bubbles of metals around halos (e.g. \citealp{Scannapieco2006,Booth2012}) or assuming a metallicity-density relation (e.g. \citealp{Keating2014}). The $\texttt{Nyx}$ code does not account for star formation or feedback processes from stars and AGN. As chemical enrichment and transport are not modeled, we adopt the halo-based approach in post-processing to paint metals onto the baryonic distribution. We use the halo catalog from $\texttt{Nyx}$ at the same redshift snapshot, where the halo catalog is generated by topologically connecting cells whose densities are 138 times the mean density \citep{Friesen2016,Sorini2018}, which gives a similar result as a friends-of-friends halo finder with a linking length of 0.168 times the mean particle separation. 

We generate a non-uniform patchy distribution of metals by assuming that halos with masses greater than a minimum mass are able to enrich their surroundings
%% SST done
%% JFH are able to enrich their surroundings (i.e. omit gases)
out to a maximum distance.
%% SST my plan is to introduce the qualitative concept here and then introduce all the notations in the next paragraph. I think including the symbols disrupt the flow somewhat.
%% JFH Introduce the symbol R here?
The 
%% SST done
%% JFH Within this radius, the gas is assumed to be uniformly enriched to a constant metallicity logZ 
gas is assumed to be uniformly enriched to a constant metallicity within the enriched regions, while no enrichment occurs outside of these regions. For simplicity, we do not sum the metallicity of the overlapping enriched regions, but rather assume they retain the constant input metallicity (this should not be a huge effect, given uncertainties on the mixing rate of the overlapping bubbles and that 45\% of our models have a metal volume-filling fraction $\geq$ 0.30).

\begin{figure*}
\centering
\includegraphics[width=0.95\textwidth]{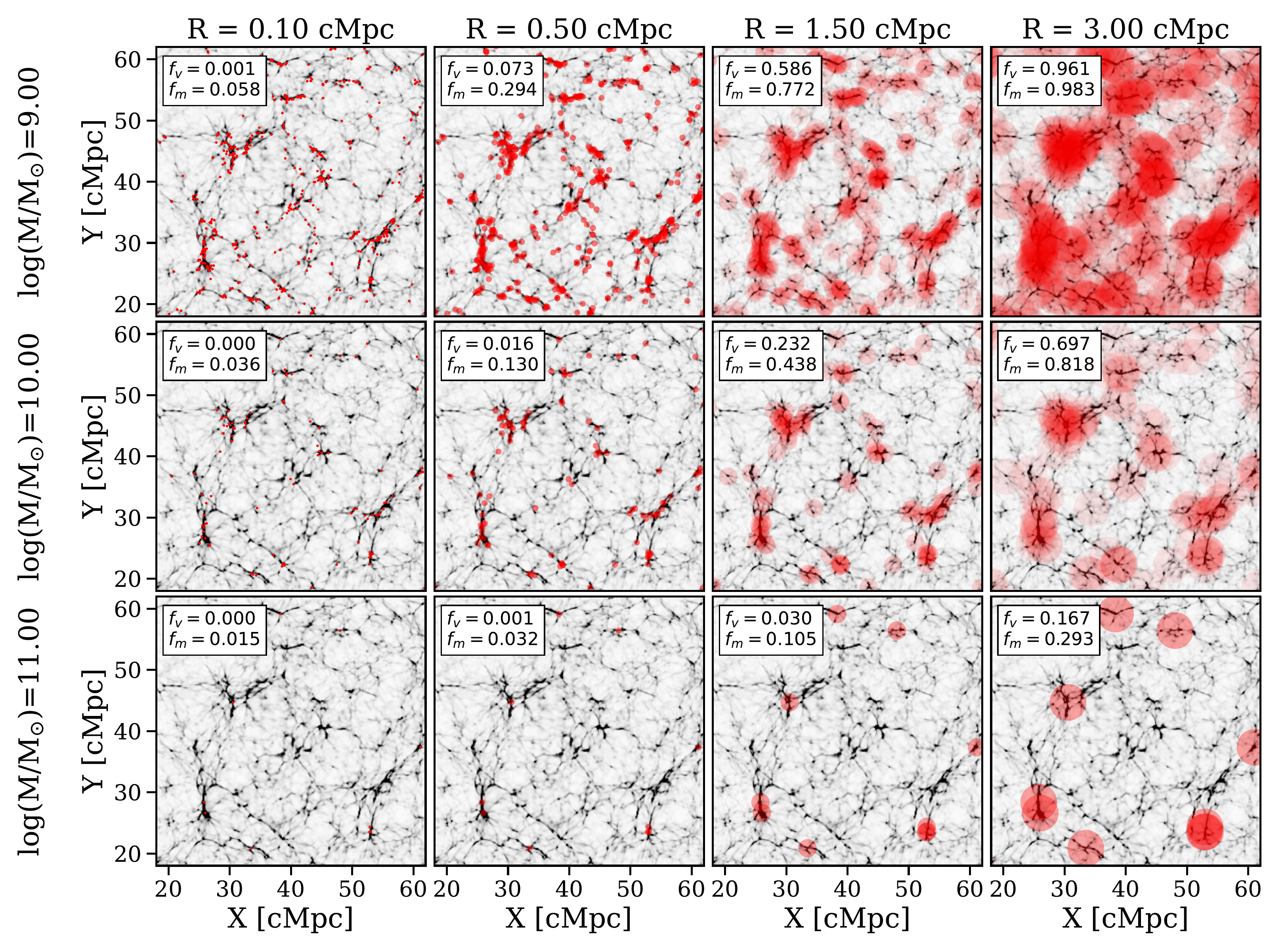}
\caption{2D slices of the Nyx density field at $z=4.5$ for different spatial distributions of metals. Each slice is 1.0 cMpc thick and shows only a portion (40 cMpc $\times$ 40 cMpc) of the entire 2D projection, and is the same for every panel. Each row corresponds to a different minimum halo mass that contributes to enrichment (log$\,M$) and each column corresponds to a different maximum radius out to which enrichment occurs ($R$). For each model, the enriched regions are shown as red shaded regions. For a fixed row, the same set of halos contribute to the underlying enrichment, but out to increasing distances from left to right. For a fixed column, the enrichment radius is fixed but different halos masses give rise to the enrichment. The corresponding volume-filling fraction ($f_V$) and mass-filling fraction ($f_m$) of metals are indicated. The filling fractions increase with decreasing log$\,M$ and increasing $R$.}
\label{enrichedhalos}
\end{figure*}

We create grids of minimum halo mass log$_{10}\,(M_{\mathrm{min}}/M_\odot)$ from 8.5 to 11.0 in increments of 0.10 and grids of maximum enrichment radius $R$ from 0.1 to 3.0 cMpc in increments of 0.1 cMpc, for a total of 780 enrichment topologies. We vary the $[\mathrm{C/H}]$ metallicity of the enriched gas from $-4.5$ to $-2.0$ in increments of 0.1. For the rest of this paper, we will sometimes use the shorthand log$\,Z$ to mean [C/H]. Figure \ref{enrichedhalos} shows our imposed metal distributions for a representative subset of models that span our parameter grid.
%% SST done
%% JFH "as a function of a subset of the models" is not proper English. I would say "for a representative set of models covering
%% our parameter grid" or something like that. 
Going horizontally across each row, the same set of halos that are above some minimum mass give rise to the IGM enrichment (red shaded regions), but out to increasing distances from left to right. Going vertically down each column, the enrichment distance is fixed but different sets of halos contribute to the enrichment. Uniform enrichment occurs in the limiting case where all halos contribute to the enrichment out to infinitely large $R$. 

\begin{figure*}
\centering
\includegraphics[width=0.9\textwidth]{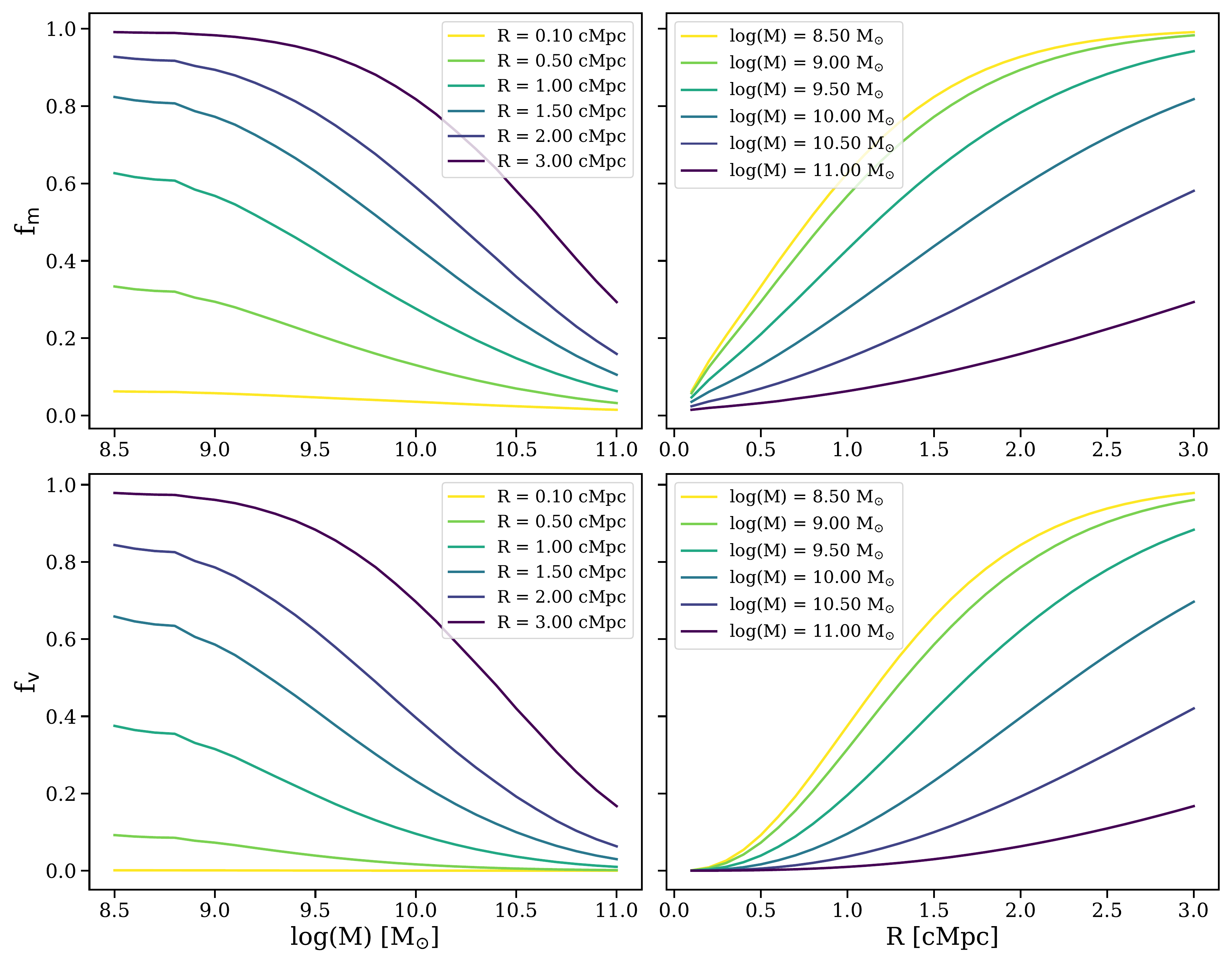}
\caption{Trends of the mass-filling fraction ($f_m$; top panel) and volume-filling fraction ($f_V$; bottom panel) as a function of the enrichment topology model parameters. The left panel shows the relation as a function of log$\,M$ at fixed $R$ (different colored lines). For very small $R$, the filling fractions remain almost constant with log$\,M$, otherwise they decrease with increasing log$\,M$ (less contributing halos). The right panel shows the relation as a function of $R$ at fixed log$\,M$ (different colored lines). The filling fractions decrease with decreasing $R$ (smaller enrichment region). For the same enrichment topology, $f_m > f_V$ because $f_m$ is weighted towards overdense region whereas $f_V$ is a simple number count. }
\label{fvfm}

%% SST mentioned in caption that the input metallicity is the maximum value on the y-axis. 
%% JFH Indicate the input metallicity as a horizontal dashed line or something like that. 
\centering
\includegraphics[width=0.9\textwidth]{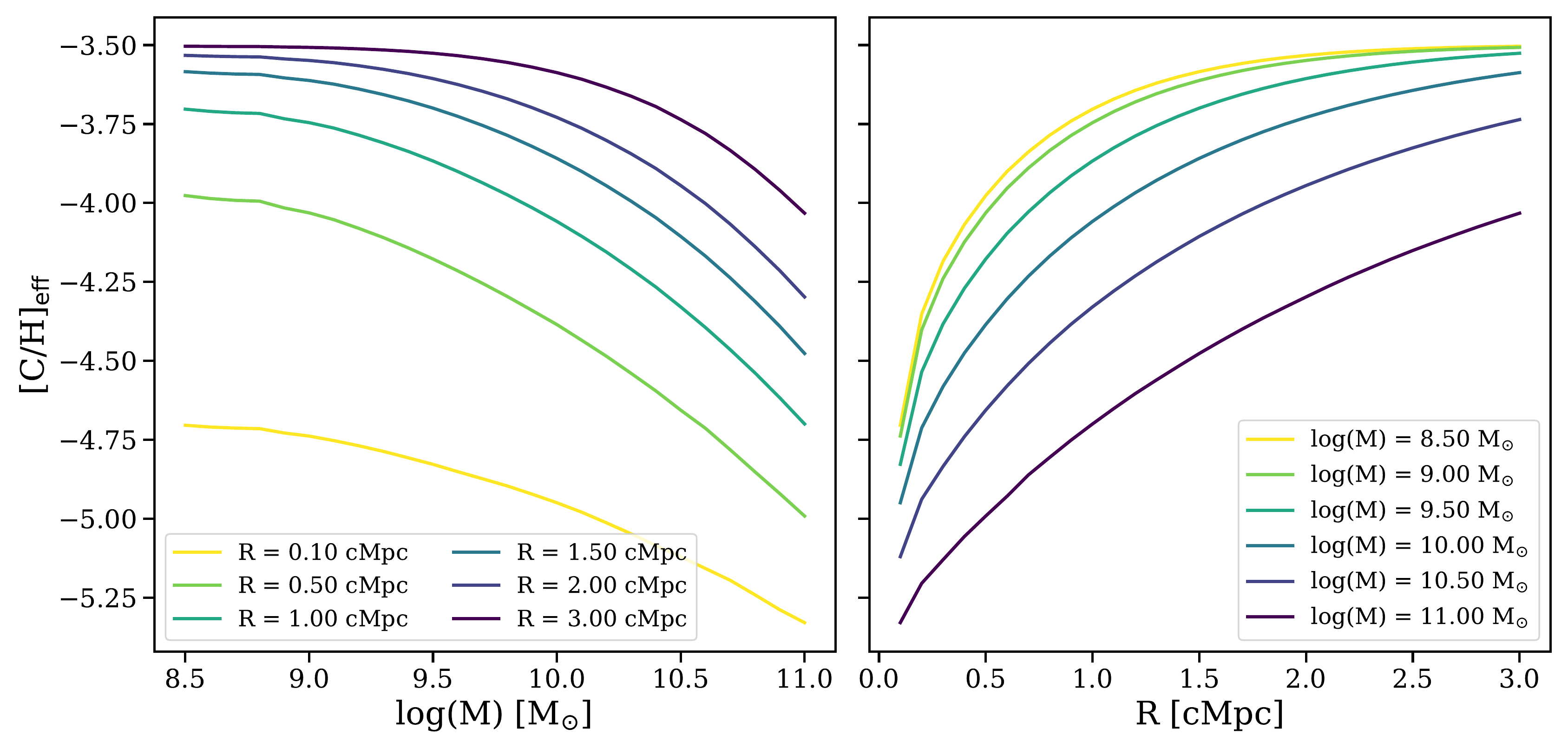}
\caption{The effective metallicity as a function of the enrichment topology model parameters, log$\,M$ (left) and $R$ (right). The maximum value on the y-axis indicates the input metallicity of [C/H] = $-3.50$. The effective metallicity reflects the underlying trend of the mass-filling fraction $f_m$, with the input metallicity affecting the overall normalization (see Eqn \ref{eqn-logzeff2}).}
\label{logZeff}
\end{figure*}

We compute the corresponding mass- and volume-filling fractions of metals from our enrichment topologies. The mass-filling fraction ($f_m$) is calculated as the total densities
%% SST done: changed to total density of all pixels
%% JFH I'm not sure what a total density is. I would say "sum of the densities of the enriched pixels over the sum"
of the enriched pixels (denoted with superscript $Z$) over the total density of all pixels. The volume-filling fraction ($f_V$) is calculated as the total number of enriched pixels over all pixels.

%% SST done: it's already defined in that paragraph
%% JFH YOu need to define the Z superscripted symbols somewhere in the text. 
\begin{align}
f_m = \frac{\Sigma_i \rho_i^Z}{\Sigma_i \rho_i} = \frac{\Sigma_i \Delta_i^Z}{\Sigma_i \Delta_i} %\; ; \Delta_i \equiv \frac{\rho_i}{\langle \rho \rangle_\mathrm{V}}, 
%% SST done
%% JFH get rid of the semi-colon in this equation and just indicate the \Delta definition in the text. It is cramped and confusing with 
%% the semi-colon in my opinion. 
\end{align}
\begin{align}
f_V = \frac{N^Z}{N_{\rm{pix}}},
\end{align}

\noindent where $\Delta_i \equiv \rho_i/\langle \rho \rangle_\mathrm{V}$. As expected, Figure \ref{enrichedhalos} shows both filling fractions increasing with increasing $R$ but decreasing as log$_{10}\,M_{\mathrm{min}}$ increases, as fewer halos contribute to the enrichment. As $f_m$ is weighted towards overdense regions, it is always larger than $f_V$ for the same topology. Additionally, for topologies with similar $f_V$, 
%% SST done
%% JFH2 Additionally, for topologies with similar $f_V$, in those with the larger f_m the metals appear more concentrated ....
the metals appear more concentrated as opposed to more uniformly-distributed in those with larger $f_m$, as overdense regions are also more clustered. 
%% SST done
%% JFH Figure 2 is impossible to read. Fonts, curves, everything is way too small. Make this fill the page instead. Likewise\
%% for Figure 3. 
Figure \ref{fvfm} shows the trends of the filling fractions with log$_{10}\,M_{\mathrm{min}}$ and $R$. 

%% SST done at the end of P1.
%% JFH2 For concreteness can you define the relation n_C = Z_sol 10^[C/H] n_H, and also quote the value you use for the 
%% solar carbon abundance. This could be done for example in the last paragraph of the intro. 
Given an inhomogeneous enrichment with input metallicity [C/H], one can also compute the \textit{effective} metallicity of the IGM  
%% SST typos, fixed
%% JFH Lower case ns are confusing here since that usually represnets a number density, not a total number. Use N_C and N_H instead. 
\begin{align}
Z_{\mathrm{eff}} 
&\equiv \frac{N_\mathrm{C}}{N_\mathrm{H}} \nonumber \\
&= \frac{Z_{\odot} 10^{\rm{[C/H]}} \langle n_{\mathrm{H}} \rangle \Sigma_i (\Delta_i^Z \cdot V_{\mathrm{cell},i})}{\langle n_\mathrm{H} \rangle \Sigma_i(\Delta_i \cdot V_{\mathrm{cell},i})} \nonumber \\
%% SST removed the following line for clarity
%% JFH The semi-colon V_box look bad and crapmped here. Just put in the text. I'm not sure you even need to introduce it since it just cancels. I'm also 
%% not following your math, i.e. introducing V_box is effectively adding another summation, but I don't see where that is coming from?? This is a lot of equations
%% to indicate an extremely simple concept and is so convolute that I cannot follow what you are doing. Plesae simplify. 
%&= \frac{Z_{\odot} 10^{\rm{[C/H]}} \langle n_\mathrm{H} \rangle V_{\rm{box}} \Sigma_i (\Delta_i^Z)}{\langle n_\mathrm{H} \rangle V_{\rm{box}} \Sigma_i(\Delta_i)} \; ; V_{\rm{box}} = \Sigma_i (v_{\mathrm{cell},i})\nonumber \\
&= Z_{\odot} \cdot 10^{\rm{[C/H]}} \cdot \frac{\Sigma_i \Delta_i^Z}{\Sigma_i \Delta_i} \nonumber \\
&= Z_{\odot} \cdot 10^{\rm{[C/H]}} \cdot f_m \nonumber  %\mathrm{ ,} 
%\label{eqn-logzeff}
\end{align}
\noindent
\begin{align}
%% SST done
%% JFH2 I think this equal should be changed to \equiv since this is a definition of C/H_eff
\mathrm{[C/H]}_{\rm{eff}} \equiv \mathrm{log}_{10}\left(\frac{Z_{\rm{eff}}}{Z_{\odot}}\right) = \mathrm{log}_{10} \left( 10^{\mathrm{[C/H]}} \cdot f_m \right),
\label{eqn-logzeff2}
\end{align}

\noindent where $N_\mathrm{C}$ is the total number of carbon atoms, $\Delta_i^Z$ is the overdensity of enriched pixel $i$, and $V_{\mathrm{cell},i}$ is the volume of a cell in the simulation. Figure \ref{logZeff} shows [C/H]$_{\rm{eff}}$ for an input [C/H] = $-3.50$ while varying the morphological parameters. 
%% SST it's already defined in the prev paragraph
%% JFH The delta^z parameter needs to be introduced earlier, i.e. in the equation above since you use it already there. 

\subsection{Creating \ion{C}{\uppercase{IV}} skewers}
\label{createskewers}
The methodology to simulate the \ion{C}{IV} forest is similar to the methods detailed in \S2.4 of \citetalias{Hennawi2020} to simulate the \ion{Mg}{II} forest. We first randomly draw \texttt{Nyx} skewers along one face of our simulation box, for a total of 10,000 skewers. These provide us with the baryon density, temperature, and line-of-sight peculiar velocity. We additionally need the fraction of carbon in the triply-ionized state for each cell of the skewer. We compute the ionization fractions of carbon at $z=4.5$ using \texttt{CLOUDY} (C17 version; \citealp{Ferland2017}) on a grid of hydrogen densities (log $n_\mathrm{H}$ = $-7$ cm$^{-3}$ to log $n_\mathrm{H}$ = 0 cm$^{-3}$ in increments of 0.1 dex), gas temperatures ($10^2$ K to $10^7$ K in increments of 0.1 dex), and metal abundance ([C/H] from $-3.5$ to $-1.5$). The gas is irradiated by a uniform UV background from galaxies and quasars using the prescription of \cite{HM2012}. As the \ion{C}{IV} fraction does not vary significantly with metal abundance (for an optically thin IGM in photoionization equilibrium), we use the results obtained with [C/H] = $-3.5$ for the rest of this paper. 

\begin{figure*}
\centering
\includegraphics[width=0.95\textwidth]{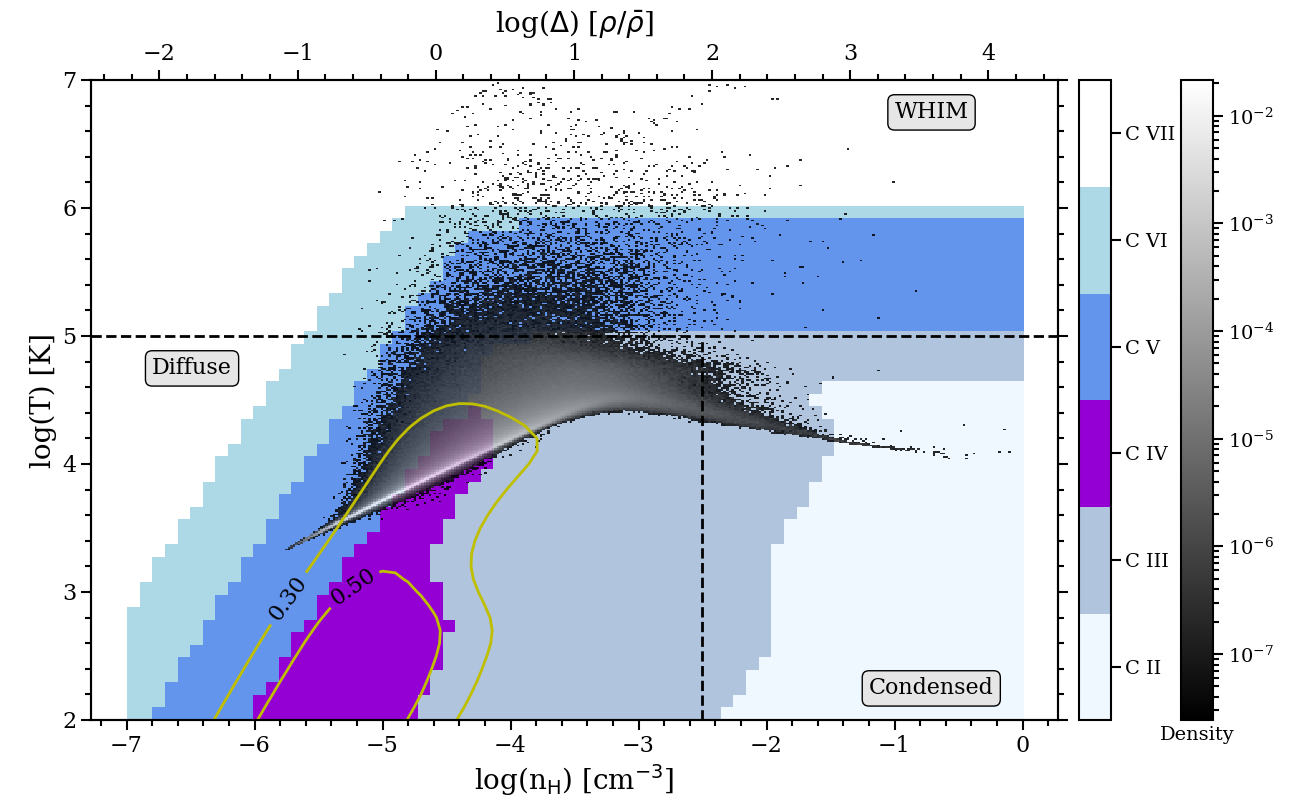}
\caption{The dominant carbon ions (indicated by colorbar) as a function of temperature and hydrogen density, overplotted against the distribution of our 10,000 random \texttt{Nyx} skewers, with the gray scale showing the density of points. The dominant ion is determined based on the ionization fraction output by \texttt{CLOUDY} at $z=4.5$. The temperature and density phase space is demarcated into ``diffuse" IGM, ``condensed" gas, and ``WHIM" (warm hot intergalactic medium) gas by dashed lines. The \ion{C}{IV} ion (purple region) dominates in the very low-density and cold IGM, and the yellow lines show the contours for \ion{C}{IV} fraction of 0.30 and 0.50. At high temperatures $> 10^5$ K, collisional ionization takes over and is denoted by the horizontal band.}
\label{rhoT-ionfrac}
\end{figure*}

Figure \ref{rhoT-ionfrac} shows the dominant carbon ions in the density and temperature phase space of our \texttt{CLOUDY} grid, overplotted against the distribution of our randomly-selected skewers. The dominant ion is defined as the ion that has the largest fraction among all available ions at each grid point (the value of this maximum fraction varies). The \ion{C}{IV} ion is the most dominant ion in the very low-density (log $n_\mathrm{H} \sim -6$ to $-5$ cm$^{-3}$) and cool
%% SST changed to gas
%% JFH2 I would change this to cool instead of cold. Also, I would say T = 100K cannot be IGM  as the plot clearly shows so you should reword or remove IGM after you say this. 
(10$^2 - 10^{4.5}$ K) gas, with an ionic fraction of $x_{\mathrm{\ion{C}{IV}}} \sim 0.45$ at mean density. The \ion{C}{II} and \ion{C}{III} ions dominate in the condensed (log $n_\mathrm{H} > -2.50$ cm$^{-3}$) and moderately low-density (log $n_\mathrm{H} \sim -4$ to $-2.50$ cm$^{-3}$) gas phases,
%% SST done
%% JFH2 Get into the habit of quoting quantitatively the regions of the plot you want people to look at. Condensed is labeled so that is fair, but moderately low-density means nothing unless you put in some numbers about where to look in the figure. 
respectively, whereas high-ionization ions typically reside in hot $T \gtrsim 10^5$ K
%% SST done
%% JFH indicate what you mean by hot, i.e. T \sim XXX?
gas in both the diffuse and condensed phases. The horizontal bands of \ion{C}{V} and \ion{C}{VI} ions
%% SST slightly clarified above
%% JFH Make it more clear what you are referring to by horizontal band
at $T \sim 10^5$ K that span log $n_\mathrm{H} > 10^{-5}$ cm$^{-3}$ are due to collisional ionization. We linearly interpolate the \texttt{CLOUDY} output of $x_{\mathrm{\ion{C}{IV}}}$ as a function of density and temperature onto the \texttt{Nyx} skewers. Given these ingredients, the optical depth for each component of the \ion{C}{IV} doublet can be computed as
%% SST done
%% JFH2 In the equation below nad other places you have tau_C etc which should be tau_{\rm C}. Element names should not be in math mode. 
\begin{align}
%% SST done
%% JFH2 I'm fine with your E,01 however the enrichment topology is in real space and tau is in redshift
%% space, so the E_0,1 term needs to be under the intgral not outside it!
%% SST Since x_civ = 0 in regions that are not enriched, why do we need the extra topology term? I get that it's can be confusing since x_civ = 0 when C is in some other ionization state, so it can zero due to ionization or zero due to enrichment. Anyway, I agree that it's best to be explicit so I inserted the E_0,1 term for topology. Not sure if this is the best variable to use.
%% JFH You are defining x_CIV to be the CIV fraction above whereas here in the equation it is actually x_CIV*x_C where x_C is the enrichment 
%% topology which one or zero. I'm not sure that x_C is the the best notation but right now you are not
%% making this explicit. It is called "enrichment topology" in Figure 5. Equation 4 as it is written is missing a term. 
    \tau_v = \tau_{\rm C,0} \int E_{0,1} \frac{x_{\mathrm{\ion{C}{IV}}}\Delta}{\sqrt{\pi}}\mathrm{exp}\left[-\left(\frac{v' - v}{b}\right)^2 \right]\frac{dv'}{b},
\label{tau}
\end{align}
\noindent where $E_{0,1}$ is the enrichment topology (in practice, a True/False bit that determines if a location is enriched or not), $b$ is the Doppler parameter and $\tau_{\rm C,0}$ is the 
%% SST done
%% JFH analog of the Gunn-Peterson optical depth (since that only refers to hydrogen)
\ion{C}{IV} analog of the Gunn-Peterson optical depth  
\begin{align}
    \tau_{\rm C,0} &= \frac{\pi e^2 f_{lu} \lambda_{lu} n_{\rm C}}{m_e c H(z)}. 
\end{align}
%% SST removed square brackets and clarified Z
%% JFH Notation is a bit confusing since it does not make sense to discuss volume averages here since you have a topology which varies. I would 
%% use n_X_0 or something and reserve volume averages for true volume averages.  I also think you need to clarify that this is the constant b/g metallicity you are using in this equation not the Z_eff. 
\noindent The $f_{lu}$ term is the oscillator strength, $\lambda_{lu}$ is the wavelength of the transition, and $n_{\rm C}$ is the number density of carbon, which is related to the metallicity as $n_{\rm C} = (Z/Z_\odot)(n_{\rm C}/n_{\rm H})_\odot \langle n_{\rm H} \rangle$. At $z=4.5$ and assuming an input [C/H] = $-3.5$ in enriched regions,
\begin{align}
    \tau_{\rm C,0} &= 0.013 \left(\frac{Z/Z_\odot}{10^{-3.5}}\right) \left(\frac{(n_{\rm C}/n_{\rm H})_\odot}{2.7 \times 10^{-4}}\right) \left(\frac{f_{lu}}{0.1899}\right) \nonumber \\
    & \qquad\qquad\; \left(\frac{\lambda_{lu}}{\mathrm{1548 \r{A}}}\right) \left(\frac{1 + z}{1+ 4.5}\right)^{3/2}.
\label{tau0}
\end{align}

\noindent Note that metallicity is a multiplicative factor in front of the optical depth, so we can use the same set of skewers to generate the \ion{C}{IV} forest at different metallicities by simply rescaling the optical depth. In practice, rather than computing the optical depth for both transitions in the \ion{C}{IV} doublet, we only compute the optical depth of the stronger (bluer) line, $\tau_{1548}$, and rescale it by the oscillator strength ratio of the two lines $f_{1550}/f_{1548}$ = 0.499 to obtain $\tau_{1550}$, which is then shifted redward equivalent to the doublet separation $dv = 498$ km/s. The total optical depth of the \ion{C}{IV} forest is $\tau_{\mathrm{\ion{C}{IV}}} = \tau_{1548} + \tau_{1550}$. 
We adopt the same approach as \citetalias{Hennawi2020} in discretizing the integral in Equation \ref{tau} to handle the possibility that our simulation native velocity grid ($dv_{\rm pix}$ = 3.16 km/s)
%% SST done
%% JFH2 Make it more clear that you mean "our simulations native velocity grid"
barely resolves the small Doppler parameter of the \ion{C}{IV} ion ($b$ = 3.71 km/s at 10,000 K).
%$b$ = 1.17 km/s at 1000 K 
%% SST recomputed for 10,000 K and reworded to 'barely resolves'
%% JFH Your gas is not at 1000 K, so I'd quote a more reasonable number here. 
%% SST changed pixel grid to dv_pix. 
%% JFH2 You just used dv above to be the doublet separation and now you are calling dv the pixel grid. I suggest that above you just write v_{CIV} = 498 km/s instead of dv. 
The optical depth of the discretized grid cell is computed as (cf. Appendix B of \citealp{Lukic2015})
\begin{align}
    \tau_v = \tau_{X,0} \sum_i \frac{x_{\mathrm{\ion{C}{IV}},i}\Delta_i}{2} \left[ \mathrm{erf}(y_{i-1/2}) -  \mathrm{erf}(y_{i+1/2}) \right].
\end{align}
%% SST done
%% JFH Please cite the appendix in Lukic for this. 

\begin{figure*}
\centering
\includegraphics[width=0.97\textwidth]{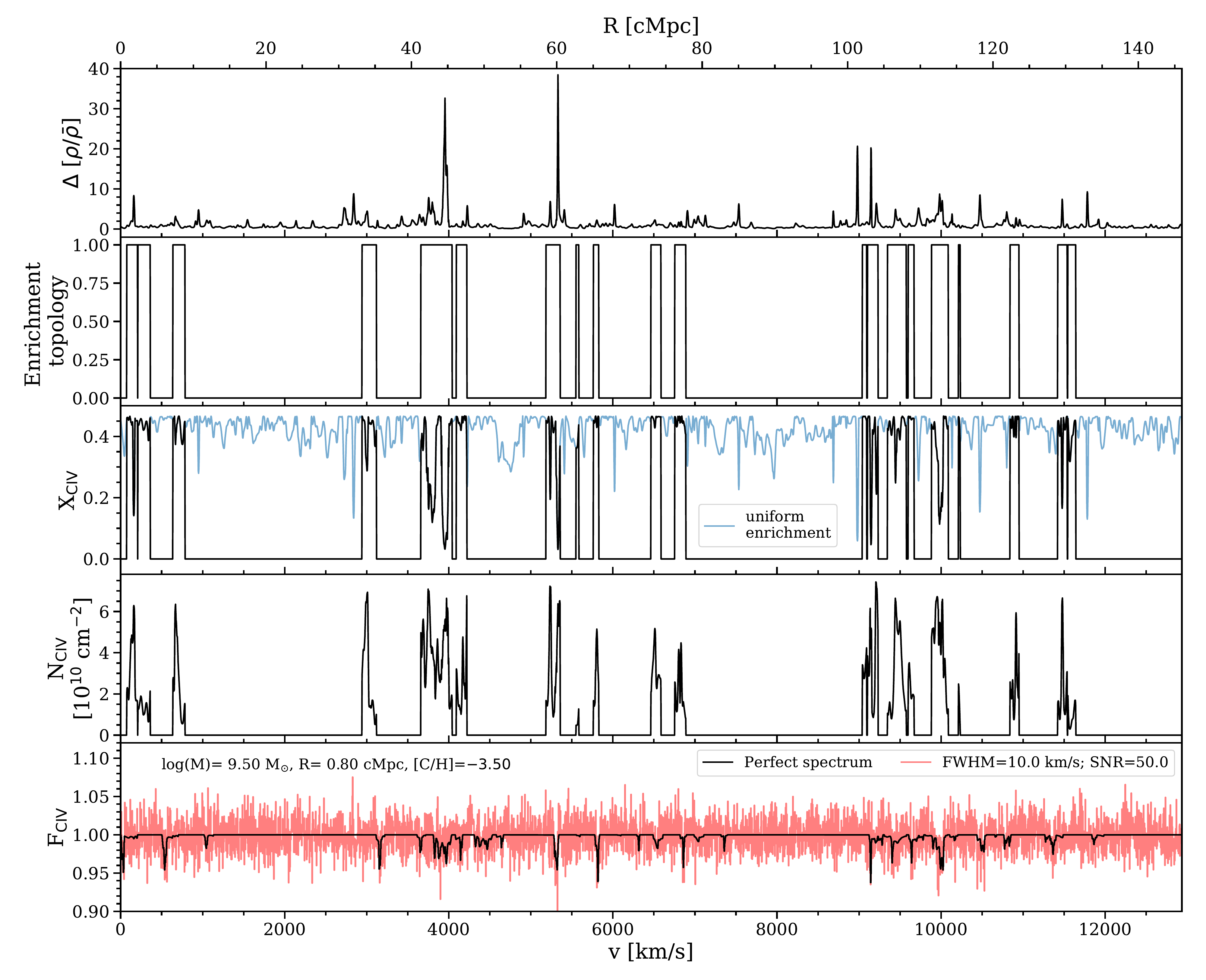}
\caption{Skewers of various quantities for a random sightline. The overdensity skewer is obtained directly from the \texttt{Nyx} box. The enrichment topology skewer is a Boolean skewer that determines which pixels are enriched (1 for enriched and 0 for not enriched), and here we show the topology for an IGM model with log$\,M$ = 9.50 $M_{\odot}$, $R$ = 0.80 Mpc, and [C/H] = $-3.50$. The corresponding \ion{C}{IV} fraction $X_{\mathrm{\ion{C}{IV}}}$  for this IGM model is shown in the middle panel, overplotted against the uniform enrichment case. We also show the column density skewer, computed as $N_{\mathrm{\ion{C}{IV}},i} = (n_{\mathrm{\ion{C}{IV}},i} \,E_{(0,1),i}) \,\times$ pixel scale, where $E_{0,1}$ is the enrichment topology. The \ion{C}{IV} forest is shown in the last panel, with black being the noiseless spectrum at the native grid resolution and red being the noisy spectrum convolved with a FWHM of 10 km/s, i.e. that of Keck/HIRES or VLT/UVES. Even with a moderately high SNR of 50 and exquisite spectral resolution, it is challenging to detect the \ion{C}{IV} forest signal.}
\label{skewers1}
\end{figure*}

Figure \ref{skewers1} shows skewers of the relevant quantities for a random sightline through our box, assuming an enrichment model with log$_{10}\,M$ = 9.50 $M_{\odot}$, $R$ = 0.80 Mpc, and [C/H] = $-3.50$. The enrichment topology skewer is essentially a Boolean skewer that determines which pixels are enriched, which subsequently determines the structures in the other skewers. We compute the \ion{C}{IV} column density for the $i$-th pixel as
%% SST included
%% JFH2 The boolean E_0,1 should enter into this equation right?
$N_{\mathrm{\ion{C}{IV}},i} = (n_{\mathrm{\ion{C}{IV}},i} \,E_{(0,1),i}) \times$pixel scale, 
%% SST done
%% JFH Note that scale again here and maybe add a note or a footnote that a typical CIV absorber would be comprised of multiple pixels. It would also be useful to 
where $n_{\mathrm{\ion{C}{IV}},i} = \Delta_i \langle n_{\rm H} \,\rangle (Z_\odot \, 10^{[\mathrm{C/H}]}) \,x_{\mathrm{\ion{C}{IV}},i}$ and our 
%% SST corrected wrong distance
%% JFH2 N_CIV is a physical quantity that involves multiplying a physical density into a proper distance. 
%% You are quoting a comoving pixel scale here which is awkward in this context. Furthermore, please
%% check that you are doing this right in the plot, i.e. converting the pixel scale to a proper distance
%% when computing the column denisty. 
pixel scale = 6.48 kpc (such that a typical absorber will span multiple pixels). 
%% SST moved
%% JFH2 this sentence on metallicity being a multiplicative factor seems out of place here. You are 
%% describing the figure already. I think this sentence rather belongs above, perhaps justr after eqn. 6. 
We show a perfect noiseless spectrum at the native grid resolution and a noisy spectrum with SNR/pix = 50 that has been degraded to the spectral resolution of Keck/HIRES or VLT/UVES, which is FWHM = 10 km/s ($R$ = 30,000). We assume the same observational setups for our mock data in \S \ref{inference}.
%% SST done
%% JFH YOu need to either state the SNR and resolution used here or forward reference to the section where you indicate these numbers. You probably should do 
%% do both. 

%% SST broke
%% JFH2 Consider a paragraph break here??
For the IGM model shown in Figure \ref{skewers1}, one can see remarkable fluctuations of order a few percent from the perfect spectrum. However, these get challenging to detect in real data, even with moderately high SNR/pix of 50 and exquisite spectral resolution. In this regime, a statistical method like the correlation function is better suited for the task. Additionally, that the column density of the diffuse IGM is $\sim 10^{10} - 10^{11}$ cm$^{-2}$ makes it extremely challenging to detect with the standard line-fitting method, even with state-of-the-art
%% SST done
%% JFH2 state-of-the-art instrumentation on the largest ground based telescopes
instrumentation on the largest ground-based telescopes. The two deepest spectra of quasars ever taken are that of B1422+231 ($z$ = 3.62; \citealp{Ellison2000}) and HE0940-1050  ($z$ = 3.09; \citealp{DOdorico2016}). The former has a SNR of $200-300$ redward of Ly$\alpha$ with a detection limit at log($N_\mathrm{\ion{C}{IV}}$/cm$^{-2}$) $\approx$ 11.6. The spectrum of HE0940-1050 has a SNR of $320-500$ in the \ion{C}{IV} forest region, where they are sensitive down to log($N_\mathrm{\ion{C}{IV}}$/cm$^{-2}$) $\approx$ 11.4. Although these high SNR observations likely probe the entire spectrum of absorbers, from the strongest CGM absorbers to the diffuse IGM absorbers at the low column density end, they are at a much lower redshift than $z=4.5$ that we simulate here. As such, nearly all detections of absorbers with $N_{\rm \ion{C}{IV}} \gtrsim 13-14$ cm$^{-2}$ at $z \gtrsim 3$
%% SST done
%% JFH2 put in a redshift z > XX and also put a column density like 10^14 cm^2. This is a rather important
%% point and you are kind of glossing over it all. I think a lot of observers things that 10^14 cm^2 absorbers come from the IGM, whereas you clearly show that these absorbers are more like 10^12
are most likely from the CGM. 

%% SST done
%% JFH Here it would be good to stress that the based on e.g. the N_CIV panel that the column densities of IGM of ~ 10^12. You could compare that to limits on CIV
%% column density from e.g. Simcoe et al. and again get across the point that the IGM absorbers are really hard to detect and that nearly all detections at this
%% this redshift are CGM. 

%% SST I think it's fine as it. I tried to truncate the plot many times previously but somehow it always messed up the plot (messing up the different plotting elements and axes etc); I eventually gave up.
%% JFH Consider reducing the range of this plot, i.e. from 3 to 7 in temperature and from -6 to -1 in density. 

\section{Results}
\label{results}
\subsection{Correlation function of the \ion{C}{\uppercase{IV}} forest}
\label{results:cf}
%* formalism and dependence on model parameters\\
To compute the correlation function of the \ion{C}{IV} forest, we first define the 
%% SST I think it's fine as is
%% JFH2 relative flux fluctuation?
flux fluctuation, 
\begin{align}
    \delta_f \equiv \frac{F - \langle F \rangle}{\langle F \rangle}, 
\end{align}
\noindent where $\langle F \rangle$ is the volume-averaged mean flux. The correlation function is then
\begin{align}
    \xi(dv) = \big\langle \delta_f(v)\delta_f(v + dv) \big\rangle, 
\end{align}
\noindent where the average is over all available pixel pairs separated by $dv$.

\begin{figure}
\centering
\includegraphics[width=0.95\columnwidth]{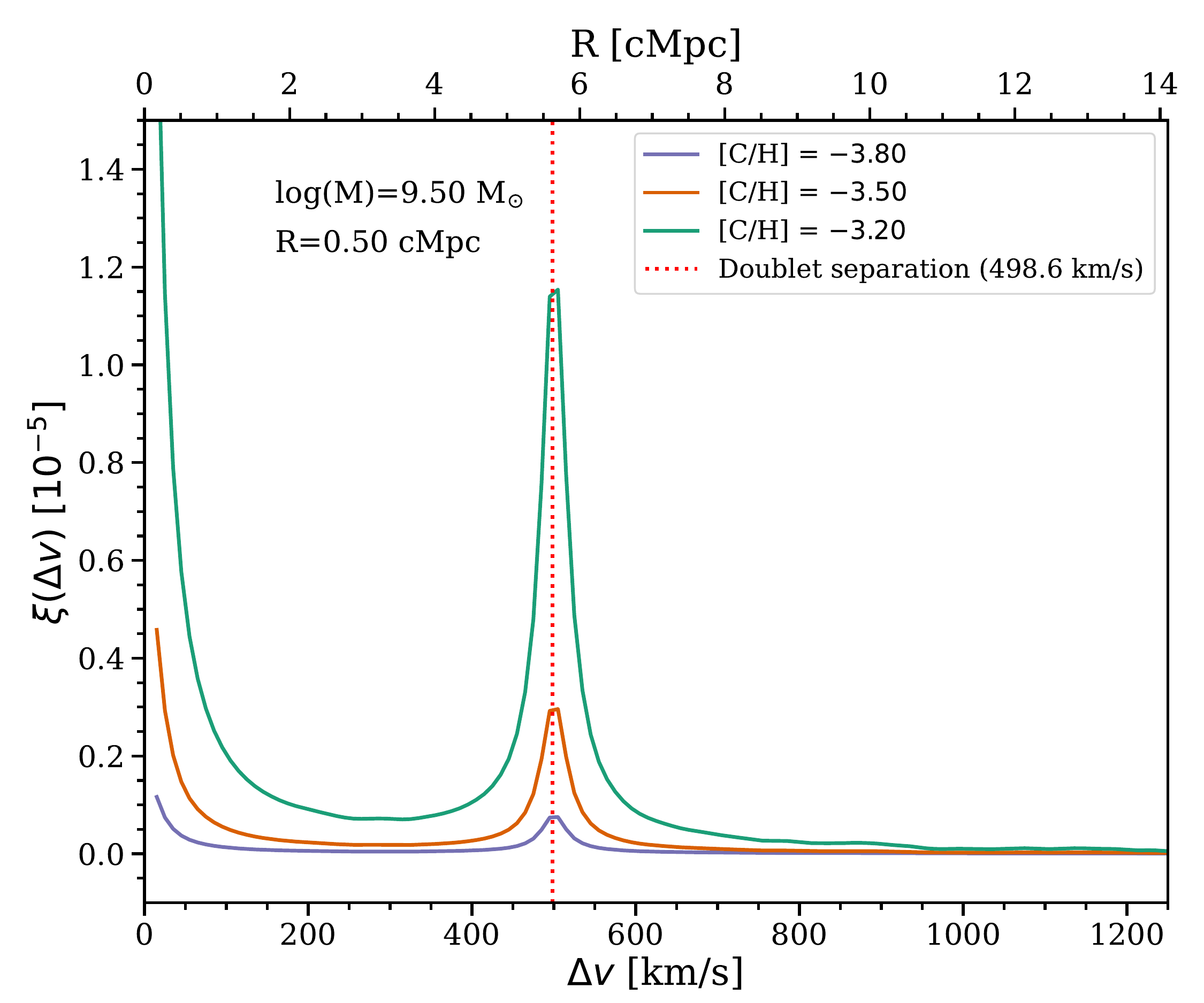}
\includegraphics[width=0.95\columnwidth]{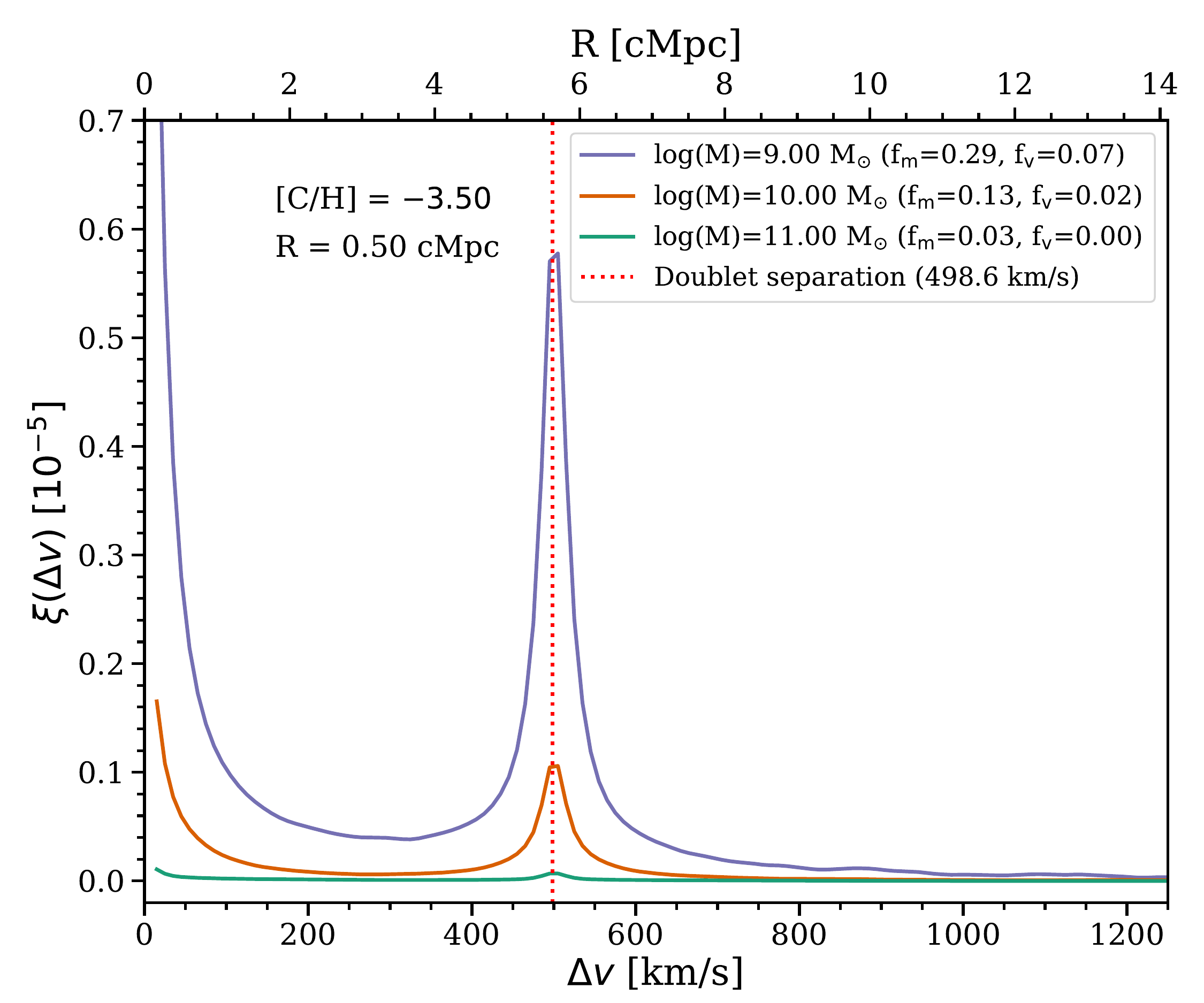}
\includegraphics[width=0.95\columnwidth]{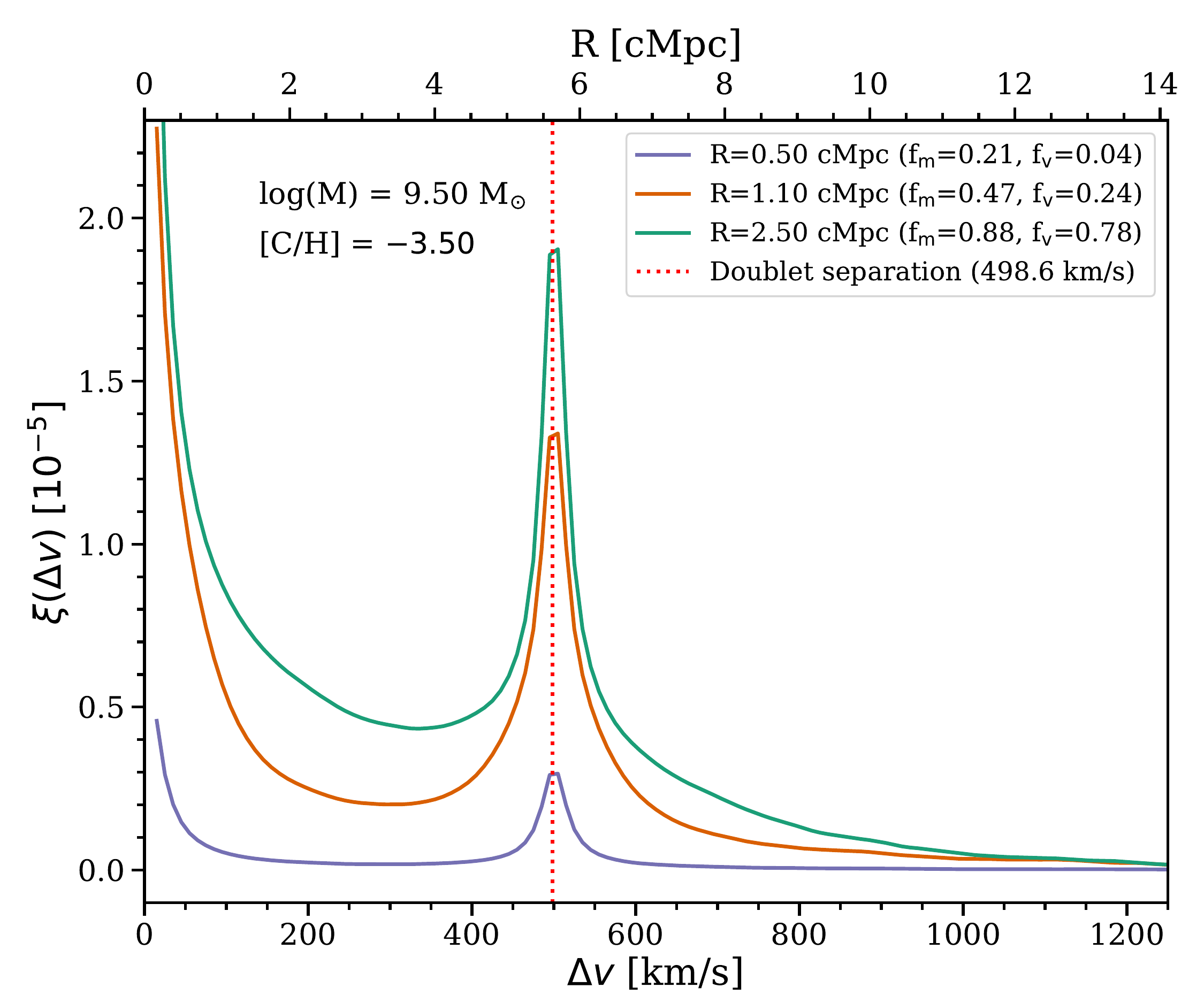}
%\caption{Correlation function of the \ion{C}{IV} forest at varying metallicity (top) and enrichment topology (bottom), computed assuming noiseless spectra convolved with FWHM = 10 km/s. The peak in the correlation corresponds to the \ion{C}{IV} doublet separation, as indicated by the vertical dotted line. The peak becomes more prominent as metallicity increases, scaling roughly as the metallicity squared. In the bottom panel, lines with the same color show models with the same log$\,M$ and lines with the same type show models with the same $R$. At fixed $R$, a larger log$\,M$ results in more large- and small-scale power on either side of the peak. The same is true for larger $R$ at fixed log$\,M$. Overall, the correlation function increases with increasing filling factor. }
\caption{Correlation function of the \ion{C}{IV} forest at varying metallicity (top) and enrichment topology (middle for varying log$_{10}M_{\mathrm{min}}$ and bottom for varying $R$), computed assuming noiseless spectra convolved with FWHM = 10 km/s. The peak in the correlation corresponds to the \ion{C}{IV} doublet separation, as indicated by the vertical dotted lines. The peak becomes more prominent as metallicity increases, scaling roughly as the metallicity squared. Increasing log$_{10}M_{\mathrm{min}}$ leads to a decrease in the peak amplitude, while increasing $R$ increases the power on both sides of the peak in the correlation function at the doublet separation.
% SST done
%%JFH2 "both sides of the peak in the correlation function at the doublet separation."
}
\label{cf_mr_logZ}
\end{figure}

Figure \ref{cf_mr_logZ} shows the correlation functions of the \ion{C}{IV} forest for varying model parameters, assuming noiseless skewers from our mock dataset (see \S \ref{inference}) that are convolved with a Gaussian line spread function of FHWM=10 km/s.
%% SST done, briefly explained here and explained in more details in the mock data section. 
%% JFH YOu never explained how skewers are resolution convolved so explain that somewhere
%% SST that's right, I just somehow set up the plot that way. I can replot but I don't think the plot will change much. Sentence reworded slightly. 
%% JFH For these noiseless plots, I don't see the point of restricting the pathlength. You are just trying to show trends in the average model. 
We see the correlation functions peaking at $dv = 498$ km/s, corresponding to the doublet separation that sets the characteristic scale of the \ion{C}{IV} forest. 

Metallicity is a normalizing factor in front of the optical depth (Equation \ref{tau0}) such that optical depth increases with metallicity and gives rise to a stronger signal. In the limit of small optical depths as is for the \ion{C}{IV} forest, $F \approx 1 - \tau \propto Z$, so metallicity affects the correlation function by rescaling the peak amplitude approximately as the square of the metallicity, e.g. the peak amplitude for [C/H] = $-3.20$ is $\sim (10^{-3.20}/10^{-3.50})^2 \sim 4$ times larger than for [C/H] = $-3.50$. 
%% SST clarified
%% JFH Okay that is true but the f = exp(-tau) so the reason for this quadratic dependence is that the optical depth is small, so you might also explain that point. 
Varying log$_{10}M_{\mathrm{min}}$ alone also affects the peak amplitude, with a weaker peak as log$_{10}M_{\mathrm{min}}$ increases. On the other hand, $R$ affects both the shape and amplitude; increasing $R$ results in an increase in power on both sides of the peak in the correlation function at the doublet separation. 
%% SST done
%% JFH2 make it more clear "both sides of the peak in the correlation function at the doublet separation." 
%% Same goes for the analogous text in the caption of the figure below. 
In general, the correlation function increases with increasing filling factors. This is essentially a metallicity effect, as increasing the filling factors results in more enrichment and higher metallicities, with $R$ additionally affecting the small- and large-scale powers. It is apparent that the effects of metallicity and enrichment topology are degenerate with each other, but we show that they can still be individually constrained to good precision in \S \ref{inference}.

%On the other hand, log($M$) and $R$ affect both the shape and amplitude. In general, the correlation function increases with increasing filling factors. This is essentially a metallicity effect, as increasing the filling factors results in more enrichment and higher metallicities. In particular, the effective metallicity is determined by $f_m$ (Equation \ref{logZeff}). Comparing correlation functions with the same $R$ but different log($M$), we see that a smaller log($M$) gives rise to more power at both large and small scales. The same is true when one compares correlation functions with the same log($M$) but different $R$. This increase in power at different scales is not seen when one varies the metallicity alone (top panel). While some models are easily distinguishable, others will require high quality data to be constrained. In practice, the effects of metallicty and enrichment topology are likely degenerate with each other, but we show that they can still be individually constrained to good precision in \S 3.3.

\begin{figure}
\centering
\includegraphics[width=\columnwidth]{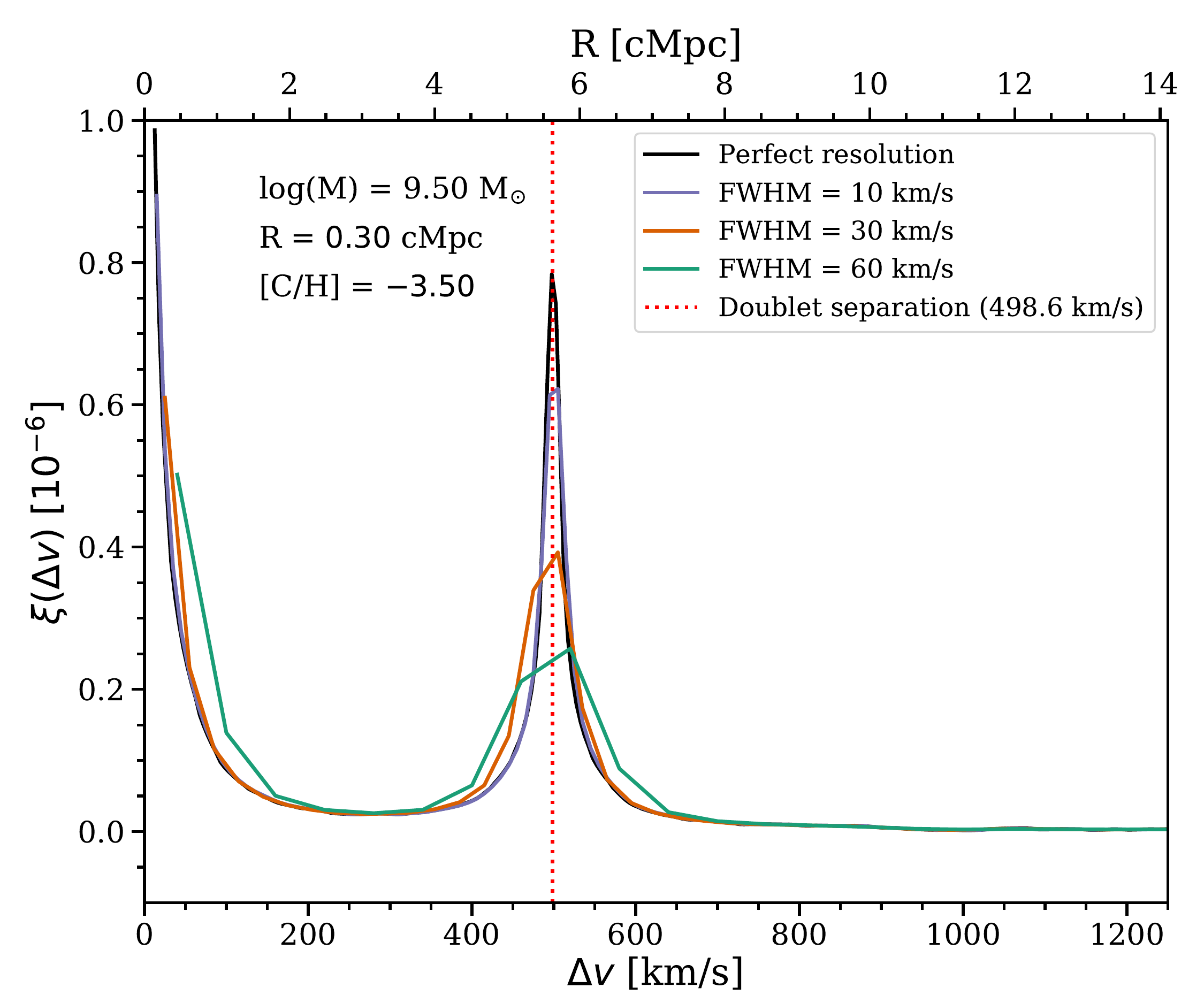}
\caption{Correlation function of the \ion{C}{IV} forest at varying spectral resolutions compared to that of a perfect spectrum (i.e. at the native resolution of our simulation), assuming noiseless spectrum. Resolutions of 10 km/s, 30 km/s, and 60 km/s are representative of Keck/HIRES (or VLT/UVES), VLT/X-SHOOTER, and Keck/DEIMOS, respectively. The particular IGM model used to compute the correlation functions is indicated. }
\label{cf_fwhm}
\end{figure}

Figure \ref{cf_fwhm} shows the effect of spectral resolution on the correlation function of the \ion{C}{IV} forest, using three resolutions that resemble Keck/HIRES and VLT/UVES (FWHM = 10 km/s), VLT/X-SHOOTER \citep{Vernet2011} (FWHM = 30 km/s), and Keck/DEIMOS (DEep Imaging Multi-Object Spectrograph; \citealp{Faber2003}) (FWHM = 60 km/s). As spectral resolution decreases, the peak of the correlation function becomes broadened and the small-scale power reduced. The peak is still visible even with FWHM = 60 km/s, although one would require high SNR data to reliably detect it. 

%% SST done
%% JFH "Contamination from CGM absorbers" as a more informative title?
\subsection{Contamination from CGM \ion{C}{\uppercase{IV}} absorbers}
\label{results:cgm}
So far our results in the previous section do not include the effect of \ion{C}{IV} absorbers from the circumgalactic medium (CGM) of galaxies, as our skewers consist only of pure IGM absorbers. As we are interested in the background metallicity of the IGM, absorbers near galaxies can bias our results as they tend to be more enriched and 
%% SST done
%% JFH resulting in larger column densities 
give rise to higher column densities (e.g. \citealp{Lehner2016,Wotta2016,Prochaska2017}). We investigate their effects here by injecting them into our skewers. 

\subsubsection{Equivalent width frequency distribution}
\label{cgmfreqdist}
We first model their abundance, given by their equivalent width frequency distribution, as a Schechter function with the following form \citep{KacprzakChurchill2011,Hasan2020}:
\begin{align}
\frac{d^2 n}{dW_{\lambda}dz} = \frac{n^*}{W*}\left(\frac{W_{\lambda}}{W^*}\right)^{\alpha} \mathrm{exp}\left(-\frac{W_{\lambda}}{W^*}\right),
\label{eqn-schechter}
\end{align}
\noindent where $n$ is the number of absorbers and $W_{\lambda}$ is the rest-frame equivalent width. 

To guide selection of the suitable model parameters, we look to existing observations. Various studies have measured the frequency distribution of \ion{C}{IV} absorbers at $z \sim 4.5$, which requires echelle spectra covering the z-band. 
%% SST done, changed to z-band.
%% JFH Not sure I'd call 8500A the NIR but maybe that is a definition. 
We are mostly interested in weak absorbers, as they are expected to be the dominant contaminating signal and because strong absorbers can be easily identified and masked. 
%% SST done
%% JFH Not sure I'd say visually since our procedur is actual automated (peak finding, masking etc.). I'd just say "strong absorbers can be easily identified and
%% and masked". 
%% SST moved and briefly summarized here. 
%% JFH2 This discussion below on Ellision and Dodorico here seems also relevant to the end of section 2 where you 
%% discuss the column   densities of IGM absorbers. Indeed, I requested that you provide more info there which you %% are providing here. Consider just moving these sentences up to end of section 2 and then you can reference
%% them again here in this discussion. 
To detect weak absorbers, the observations typically need to be taken with high-resolution and/or high SNR. The two deepest quasar spectra to date are that of B1422+231 ($z$ = 3.62; \citealp{Ellison2000}), with a detection limit of log($N_\mathrm{\ion{C}{IV}}$/cm$^{-2}$) $\approx$ 11.6, and HE0940-1050  ($z$ = 3.09; \citealp{DOdorico2016}), where they are sensitive down to log($N_\mathrm{\ion{C}{IV}}$/cm$^{-2}$) $\approx$ 11.4. Despite being very high SNR observations with the possibility of probing actual diffuse IGM absorbers, they are at a lower redshift than $z=4.5$ that we simulate here. 

Higher redshift observations exist, but at lower SNR. \cite{DOdorico2013} observed six quasars at $4.35 < z < 6.2$ with VLT/X-shooter and computed the column density distribution function (CDDF; $d^2 n/dN/dX$) in two redshift bins over the range $12.6 < \mathrm{log}(N_{\mathrm{\ion{C}{IV}}}/\rm{cm}^{-2}) < 15$. At $z < 5.3$, they are 85\% complete down to log($N_\mathrm{\ion{C}{IV}}$/cm$^{-2}$) = 13.3. \cite{Simceo2011a} measured the CDDF at $z=4.25$ with three quasar spectra obtained with the Magellan MIKE spectrograph. Their detection limit is roughly log($N_\mathrm{\ion{C}{IV}}$/cm$^{-2}$) = 12, while being substantially complete between log($N_\mathrm{\ion{C}{IV}}$/cm$^{-2}$) = 13 and 14. At the high column density end, \cite{Cooksey2013} measured the equivalent width distribution of log($N_\mathrm{\ion{C}{IV}}$/cm$^{-2}$) $>$ 14 (corresponding to $W_{\lambda} >$ 0.6 $\mathrm{\AA}$) absorbers  at $2.97 \leq z \leq 4.54$ with SDSS DR7. Their 50\% completeness limit is log($N_\mathrm{\ion{C}{IV}}$/cm$^{-2}$) = 14. One of the most comprehensive abundance measurements of \ion{C}{IV} absorbers is from \cite{Hasan2020}, who measured the equivalent width distribution of \ion{C}{IV} absorbers from 1.1 $\leq z \leq$ 4.75 using 369 QSO spectra from Keck/HIRES and VLT/UVES. Their measurements in the highest redshift bin, 2.5 $\leq z \leq$ 4.75, are 50\% complete for weak absorbers with $W_{\lambda}$ = 0.06 $\mathrm{\AA}$ (or $N_\mathrm{\ion{C}{IV}} = 1.5 \times 10^{13}$ cm$^{-2}$, see Eqn \ref{linear-cog}).
%% SST done, added Eqn 13 and 14.
%% JFH Express this also as a column density. Also, somewhere you should put in the column density equivalent width conversion ratio assuming the linear COG. 

%% SST done
%% JFH It is unclear here whether you mean N_CIV vs W_CIV or if you mean dN/dX vs dN/dz or both. Always be specific, and avoid these kinds of ambiguous statements.
We convert literature measurements, usually expressed in units of $d^2n/dN/dX$, 
% SST don't remember; removed that for noe
%% JFH2 I'm not sure why you say dX/dz here, maybe you mean  $d^2n/dN/dz$???
to be consistent with the units of our model function (Eqn \ref{eqn-schechter}), 
%% SST done
%% JFH2 Add a refernce to eqn. 10 after "model function"
via the following conversion
\begin{align}
    \frac{d^2 n}{dW_{\lambda}dz} = \frac{d^2 n}{dN dX}\left(\frac{dW}{dN}\right)^{-1}\frac{dX}{dz}. 
\end{align}
The redshift absorption pathlength $dX$ is defined as $X(z') = \int^z_{0} (1+z)^2 [H_0/H(z)] dz$ \citep{Bahcall&Peebles1969}. Assuming $\Lambda$CDM, $dX/dz = (1+z)^2 [\Omega_\mathrm{m}(1+z)^3 + \Omega_{\Lambda}]^{-1/2}$. We obtain $dW/dN$ numerically. While $W$ scales linearly with $N$ on the linear part of the curve-of-growth, 
%% SST done
%% JFH2 While $W$ scales linearly with $N$ on the linear part of the curve-of-growth, this dependence
%% changes once absorbers are saturated and we thus need to make assumptions about b values to determine
%% W(N) and hence dW/dW. Then transition to what follows
this dependence changes once absorbers are saturated (i.e. reach a high enough column density) and so one needs to assume the $b$-values in order to determine $W(N)$ and $dW/dN$. Following \citetalias{Hennawi2020}, we model the $b$-value as a sigmoid function,
%\begin{align}
\begin{multline}
b = b_{\mathrm{weak}} + (b_{\mathrm{strong}} - b_{\mathrm{weak}}) \times \\
\left[ 1 + \mathrm{exp}\left(-\frac{\mathrm{log\,}N_{\mathrm{\ion{C}{IV}}} - \mathrm{log\,}N_{\mathrm{strong}}}{\Delta\mathrm{log\,}N} \right)\right]^{-1}.
\end{multline}
%\end{align}
The $\Delta\mathrm{log\,}N$ variable is the interval over which log($N_\mathrm{\ion{C}{IV}}$) transitions from log$\,N_\mathrm{weak}$ to log$\,N_\mathrm{strong}$. For our model, we use $b_{\mathrm{weak}} = 10$ km/s, $b_{\mathrm{strong}} = 150$ km/s, log$N_\mathrm{strong}$ = 14.5, and $\Delta\mathrm{log\,}N$ = 0.35. Our values are motivated by the fact that the \ion{C}{IV} line becomes saturated at $W_{1548}$ = 0.6 $\mathrm{\AA}$ \citep{Cooksey2013}, which translates to $N_\mathrm{\ion{C}{IV}}$ = 1.5 $\times$ 10$^{14}$ cm$^{-2}$ assuming one is on the linear part of the curve-of-growth where
%% SST done
%% JFH assumign one is on the linear part of the curve-of-growth
\begin{align}
    W_{\lambda} = 0.04 \,\mathrm{\AA} \left(\frac{N_\mathrm{\ion{C}{IV}}}{10^{13}\,\rm{ cm}^{-2}}\right).
\label{linear-cog}
\end{align}
\noindent The maximum optical depth at the line center of a Voigt profile is
\begin{align}
    \tau_0 = 1.47 \left(\frac{N_\mathrm{\ion{C}{IV}}}{10^{14}\,\rm{ cm}^{-2}}\right) \left(\frac{30 \,\mathrm{km/s}}{b}\right).
\label{linear-tau}
\end{align}
Absorption lines thus 
%% SST changed
%% JFH2 COG does not saturate -- absorption saturates
saturate around $N_\mathrm{\ion{C}{IV}}$ = 1.5 $\times$ 10$^{14}$ cm$^{-2}$ ($W_{\lambda} = 0.6$ $\mathrm{\AA}$) for $b$ = 30 km/s.
%log($N_\mathrm{\ion{C}{IV}}$/cm$^{-2}$) $\sim$ 14 for a $b$-value of $\sim 65-66$ km/s. 

\begin{figure}
\centering
\includegraphics[width=\columnwidth]{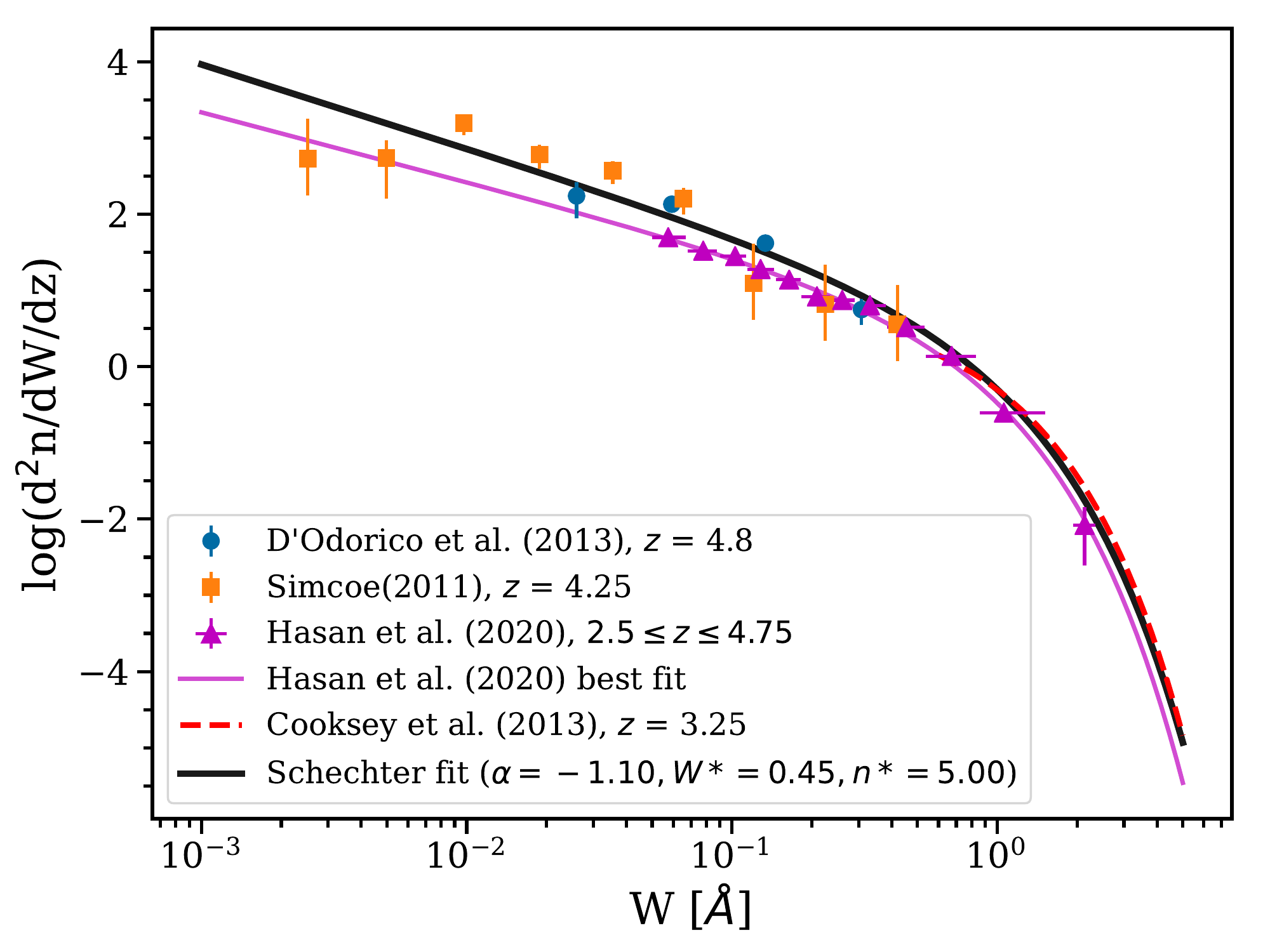}
\caption{Number of \ion{C}{IV} absorbers per unit equivalent width per unit redshift (equivalent width frequency distribution) as a function of rest-frame equivalent width. Plotted as various points are existing (high SNR and high resolution) observations at comparable redshifts as ours. We show our best-fit CGM model as the solid black line, which is consistent with existing observations and best-fits to observations (purple and red lines).}
\label{cgm_model}
\end{figure}

Our final CGM model is shown in Figure \ref{cgm_model}. We fine-tune the weak-end of the model to match with data from \cite{Simceo2011a}\footnote{The data points in \cite{Simceo2011a} were computed with a different cosmology in order to compare with an older work, so we obtain the updated and cosmology-corrected values from \cite{Bosman2017}.} and \cite{DOdorico2013} and the strong-end of the model to match with the best-fit from \cite{Cooksey2013}, resulting in ($\alpha$, $W^*$, $n^*$) = ($-1.10$, 0.45, 5). Our Schechter-function fit is slightly steeper than that of \cite{Hasan2020}, otherwise the parameters are generally similar. We randomly draw absorbers from our CGM model ranging from $W_{\lambda,\mathrm{min}} = 0.001$ $\mathrm{\AA}$ to $W_{\lambda,\mathrm{max}} = 5.0$ $\mathrm{ \AA}$ and artificially inject them into our mock skewers. We ignore absorber clustering and place absorbers at random velocity locations along our skewers. 

%% SST done
%% JFH2 I found compraing Omega_CIV to Omega_C sort of confusing. Does anyone actually compute Omega_CIV. Seems
%% like they should since this is the actual observable, but Omega_C is maybe more physically meaningful. If you 
%% cannot just compare Omega_CIV to Omega_CIV, then maybe quote a typical ionization correction or something here. 
%% Anyway, I would just add a sentence here maybe speculating on why we get a larger value that what is observed. 
Following the procedures in Appendix A of \citetalias{Hennawi2020}, the CGM absorbers of our model give [\ion{C}{IV}/H] = $-5.71$ (with solar values from \citealp{Asplund2009}), which corresponds to a cosmological mass fraction of $\Omega_{\ion{C}{IV}} = 3.97 \times 10^{-8}$. Assuming a \ion{C}{IV} to C fraction of 0.5 (which is reasonable given Figure \ref{rhoT-ionfrac}), we obtain [C/H] = $-5.41$ and $\Omega_{\rm C} = 7.94 \times 10^{-8}$. As a comparison, \cite{Simceo2011a} obtained $\Omega_{\mathrm{C}} = 2.7 \times 10^{-8}$ at $z = 4.3$ while \cite{Schaye2003} obtained $\Omega_{\mathrm{C}} = 2.21 \times 10^{-7}$ for gas with log($\Delta$) = $0.5 - 2.0$ (see Figure \ref{rhoT-ionfrac}) at $z = 3$. Potential reasons that could lead to our $\Omega_{\rm C}$ being $\sim 3\times$ higher than that obtained by \cite{Simceo2011a} are our CGM model including very strong absorbers up to $W_{\lambda} = 5 \,\mathrm{\AA}$ as well as having a steeper slope at the lowest column density end. 
%% SST maybe combination of our CGM model integrating up to Wmax = 5 A (whereas Simcoe doesn't include these very strong absorbers) + our steeper slope at the lowest column density end producing more weak absorbers (steeper compared to his last 2 data points) shifts everything up...
%% JFH Seems odd that we exceed Simcoe when considering only the CIV whereas he computed the total C. I think other authors (and maybe Simcoe also computed 
%% the Omega_CIV. Is the reason that we exceed them becuase you integrated down to low EW whereas observations have a cutoff? I doubt that is a big effect since 
%% Omega_CIV is column density weighted and as such is dominated by the high column density end. 

\subsubsection{Flux probability distribution function}
\label{fluxpdf}
\begin{figure}
\centering
\includegraphics[width=\columnwidth]{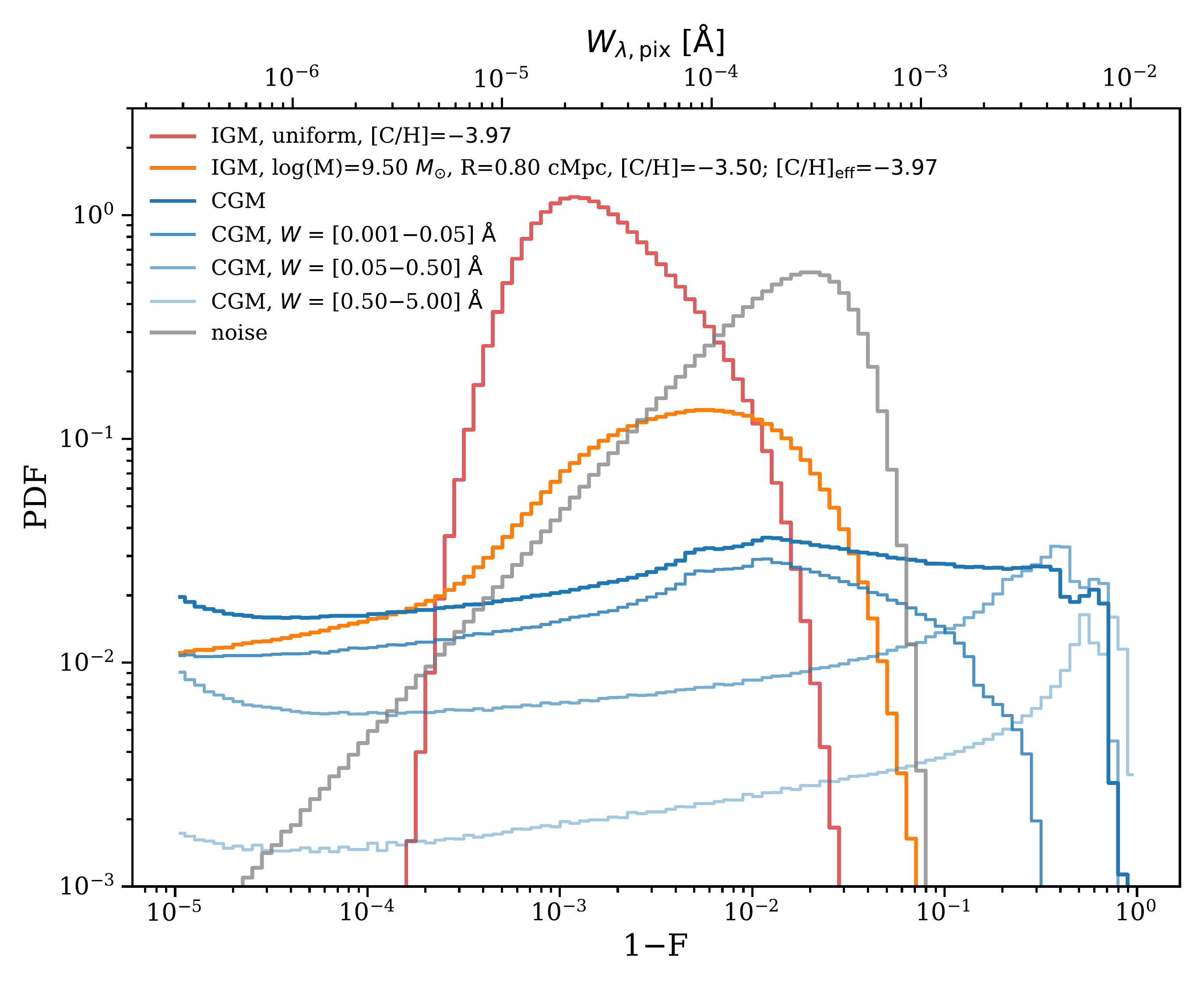}
\caption{Probability distribution function of the flux decrement for a uniformly-enriched IGM (red), a non-uniformly enriched IGM for one particular model (orange), the CGM (blue), and random noise with SNR/pix = 50 (gray). The uniform IGM model has the same metallicity as the effective metallicity of the non-uniform IGM. The CGM PDF is broken down into different equivalent width ranges, shown by the different shades of blue lines. The IGM PDFs have a distinct shape compared to the CGM PDF. They peak about small $1-F$ values, dominating over the CGM by at least a few times, and then cut off at large $1-F$ values, while the CGM PDF remains mostly flat across the entire range. }
\label{pdf_wrange}
\end{figure}

We define the flux (decrement) probability distribution function (PDF) in log$_{10}$-unit, where $\mathrm{frac}(a,b) = \int_a^b  \frac{dP}{d\,\mathrm{log}_{10}(1-F)} d\,\mathrm{log}_{10}(1-F)$ and frac($a,b$) is the fraction of pixels between $a$ and $b$. Figure \ref{pdf_wrange} shows the flux PDF of the IGM, 
%% SST done: pure IGM ---> IGM
%% JFH pure IGM is ambiguous given that you show two IGM models here, i.e. uniform and topology. 
the CGM, and Gaussian random noise with $\sigma = $ (SNR)$^{-1}$ and SNR = 50 per pixel.
%% SST done, clarified above sentence slightly and added more details in mock dataset section (3.3). 
%% JFH You never explained your noise model anywhere. Is it per pixel, per reosolution element. Is this heteroscedastci or homoscedastic noise?
The top axis of the PDF plot is $W_{\lambda,\mathrm{pix}} \equiv (1-F)\,d\lambda$, where $d\lambda$ is the width of a spectral pixel; for our mock dataset with FWHM=10 km/s, $d\lambda = 0.017 \mathrm{\AA}$ at \ion{C}{IV} 1548 $\mathrm{\AA}$. The PDFs for a uniformly-enriched IGM and for an inhomogeneously-enriched IGM with the same model as Figure \ref{skewers1} are shown, where the effective metallicity of the non-uniform IGM matches the metallicity of the uniform IGM. The PDF for the CGM component is further broken down into different $W_{\lambda}$ ranges to show the impact of absorbers with different strengths. Since the flux PDF is plotted on a log scale, negative fluctuations of the noise PDF are not shown, but they are simply symmetrical to the positive fluctuations about zero.

%% SST Done, new plot (the uniform IGM pdf is shifted a bit to the left, since changing the logZ simply shifts the pdf horizontally)
%% JFH On the legend label the metallicity of the uniform IGM model as well. Also quote the C/H eff for the non-uniform IGM. I think it might make more sense to 
%% to show a uniform model with logZ equal to the logZ_eff of the non-uniform model. That better illustrates the shape difference. 

Compared to the IGM flux PDFs that peak at small $1-F$ values and drop off at large $1-F$, the PDF for the CGM absorbers remains mostly flat up to large $1-F$ values. Weak absorbers dominate the PDF in the region of overlap with the IGM absorbers, followed by moderate and strong absorbers. As such, they are the main contaminant of our correlation function analysis. Since weak absorbers cannot produce large flux decrements, their PDF drops off at large $1-F$ values, in contrast to strong absorbers whose PDF rises and peaks at very large $1-F$ values of $\sim 1$. 

In contrast to the uniformy-enriched IGM, the PDF of the inhomogeneously-enriched IGM has a more rounded peak and a tail towards very small $1-F$ values. The rounded peak is possibly an effect of pixels being enriched pre-dominantly around halos, 
%% SST: that is a bit of circular observationally-driven statement; 
%% JFH It is not obvious to me why halos round the peak. Is that obvious to you?
while the very small flux decrements arise from mixing of metal-enriched with metal-free regions. As absorption occurs in redshift space, the optical depth of a given pixel can arise from multiple gas elements due to redshift-space distortion coming from the gas peculiar velocities. The mixing between metal-free pixels and metal-enriched pixels therefore results in this tail of very transmissive values. The flux PDF for the uniformly-enriched IGM cuts off at a value that corresponds to the most transmissive pixel (at small $F$, the flux decrement $1 - F \approx \tau$) and does not have a tail because there is no completely metal-free region. Its PDF dominates over that of the CGM by more than an order of magnitude and that of noise by a factor of a few.

\begin{figure*}
\centering
\includegraphics[width=0.97\textwidth]{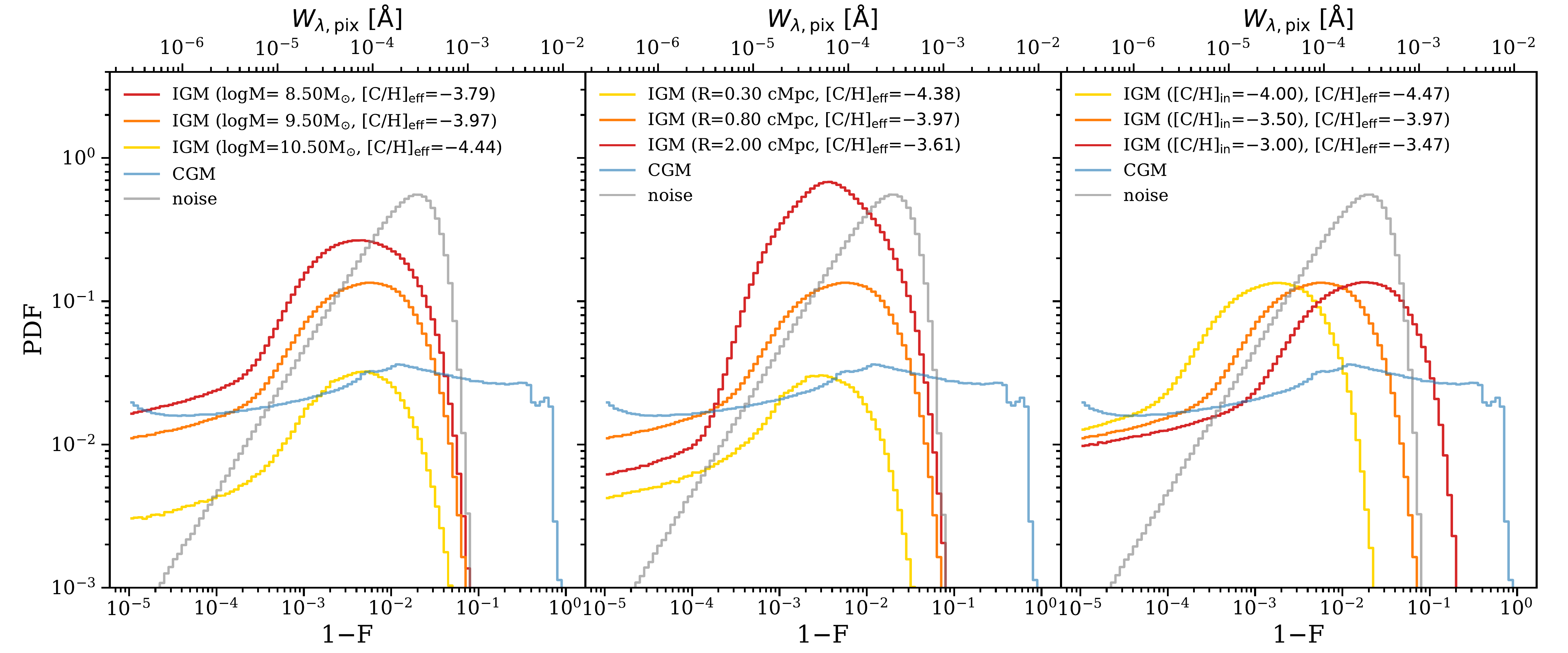}
\caption{Effects on the IGM flux decrement PDF at varying values of log$\,M$ (left), $R$ (middle), and input metallicty [C/H]$_{\mathrm{in}}$ (right). When log$\,M$ ($R$) is varied, we fix $R$ = 0.80 cMpc (log$\,M$ = 9.50 $M_{\odot}$) and [C/H]$_{\mathrm{in}} = -3.50$. When [C/H]$_{\mathrm{in}}$ is varied, we fix log$\,M$ = 9.50 $M_{\odot}$ and $R$ = 0.80 cMpc. The CGM and noise PDFs are the same in all panels. Increasing log$\,M$ shifts the PDF downward, while increasing $R$ has the opposite effect. Increasing [C/H]$_{\mathrm{in}}$ results in more absorbent pixels and shifts the PDF rightward.}
\label{pdf_mrz}
\end{figure*}

%% SST ok, changed
%% JFH2 I think Figure 10 would be clearer if you always plotted in red the model with the highest C/H_eff
Figure \ref{pdf_mrz} shows the changes in the flux decrement PDF of the inhomogeneously-enriched IGM with model parameters. From left to right, we vary log$\,M$, $R$, and [C/H] while holding the other two parameters fixed. Increasing log$\,M$ amounts to renormalizing the entire PDF downward --- because massive halos are rare, fewer pixels are enriched when restricting to higher mass halos.  
%% SST done, new plot
%% JFH Why not show a third curve in Figure 10 left for logM = 8.5? That makes it similar to the other plots. 
If only the most massive halos (log$\,M \geq$ 10.5 $M_\odot$; yellow line) contribute to enrichment, the CGM absorbers will start to be more abundant than their IGM counterparts. Increasing the enrichment radius $R$ has a similar effect as decreasing log$\,M$, resulting in a larger number of enriched pixels (the metal filling fractions approach unity as $R$ increases, see Figure \ref{fvfm}). This can be seen as the entire PDF being shifted upward. It also asymptotes 
%% SST done
%% JFH Maybe cite Figure 3 here that indicates that for the largest R the f_M approaches unity. 
to the flux PDF of a uniformly-enriched IGM for very large $R$ where the metal filling fractions approaching unity, see Figure \ref{fvfm}), with the peak of the PDF becoming more pronounced (less biased enrichment) and the tail of the distribution dropping rapidly for very small $1-F$ values (smaller number of very transmissive pixels). 
Changing [C/H] simply shifts the PDF left or right, which can be understood because optical depth, hence flux in the low optical depth limit,
%% SST done
%% JFH (and hence flux in the low optical depth limit) 
scales linearly with metallicity. A higher metallicity produces a higher optical depth that then shifts the entire PDF to the right, and vice versa. The importance of noise depends on the IGM model.  Generally, the flux fluctuations of models with smaller log$\,M$ and larger $R$ (i.e. larger [C/H$_{\rm eff}$) dominate over noise fluctuations at the peak of the IGM PDF. Nevertheless, the IGM peak may be shifted closer to or farther from that of noise depending on the input metallicity.

\subsubsection{Masking CGM absorbers}
\label{maskingcgm}
\begin{figure*}
\centering
\includegraphics[width=\columnwidth]{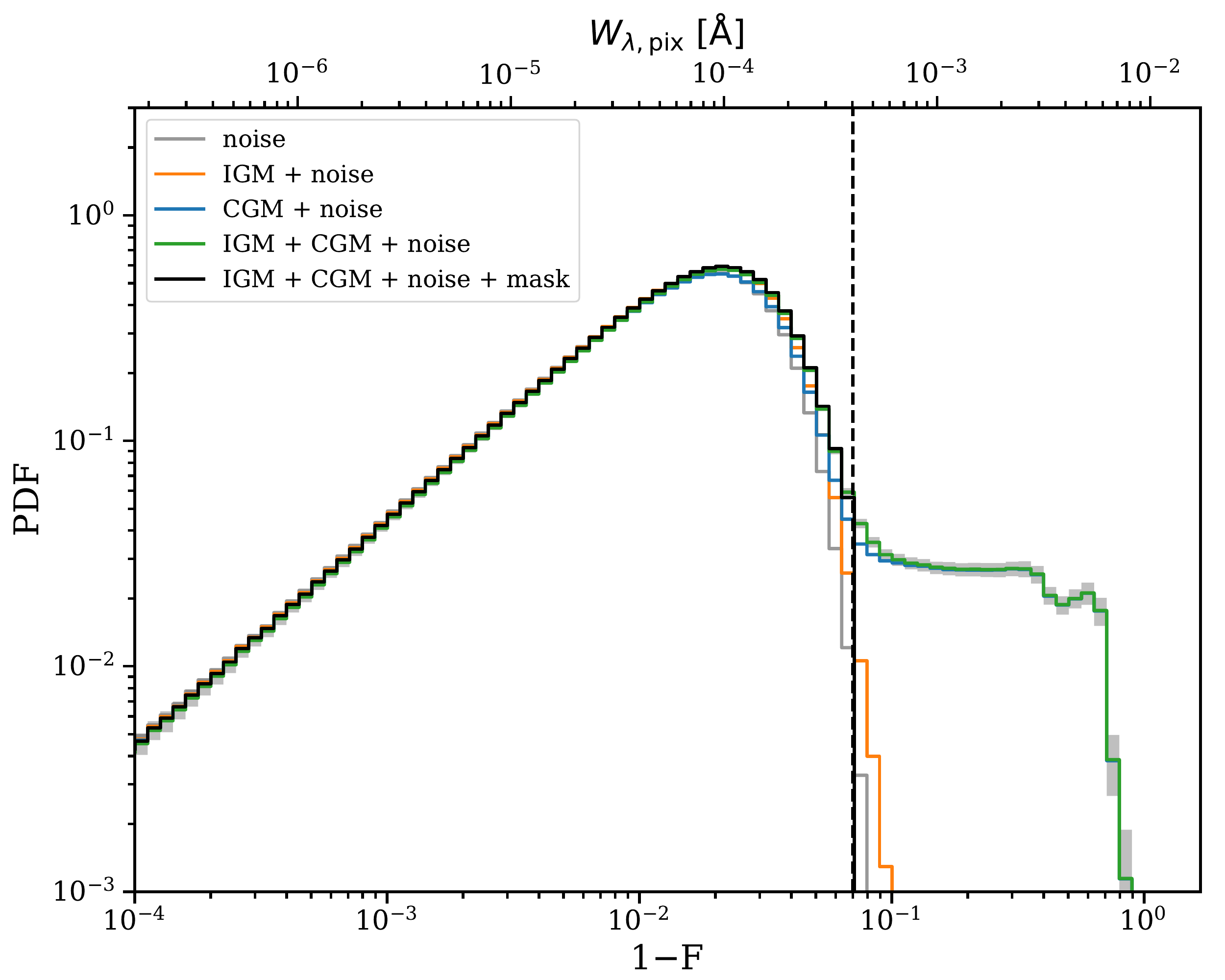}
\includegraphics[width=\columnwidth]{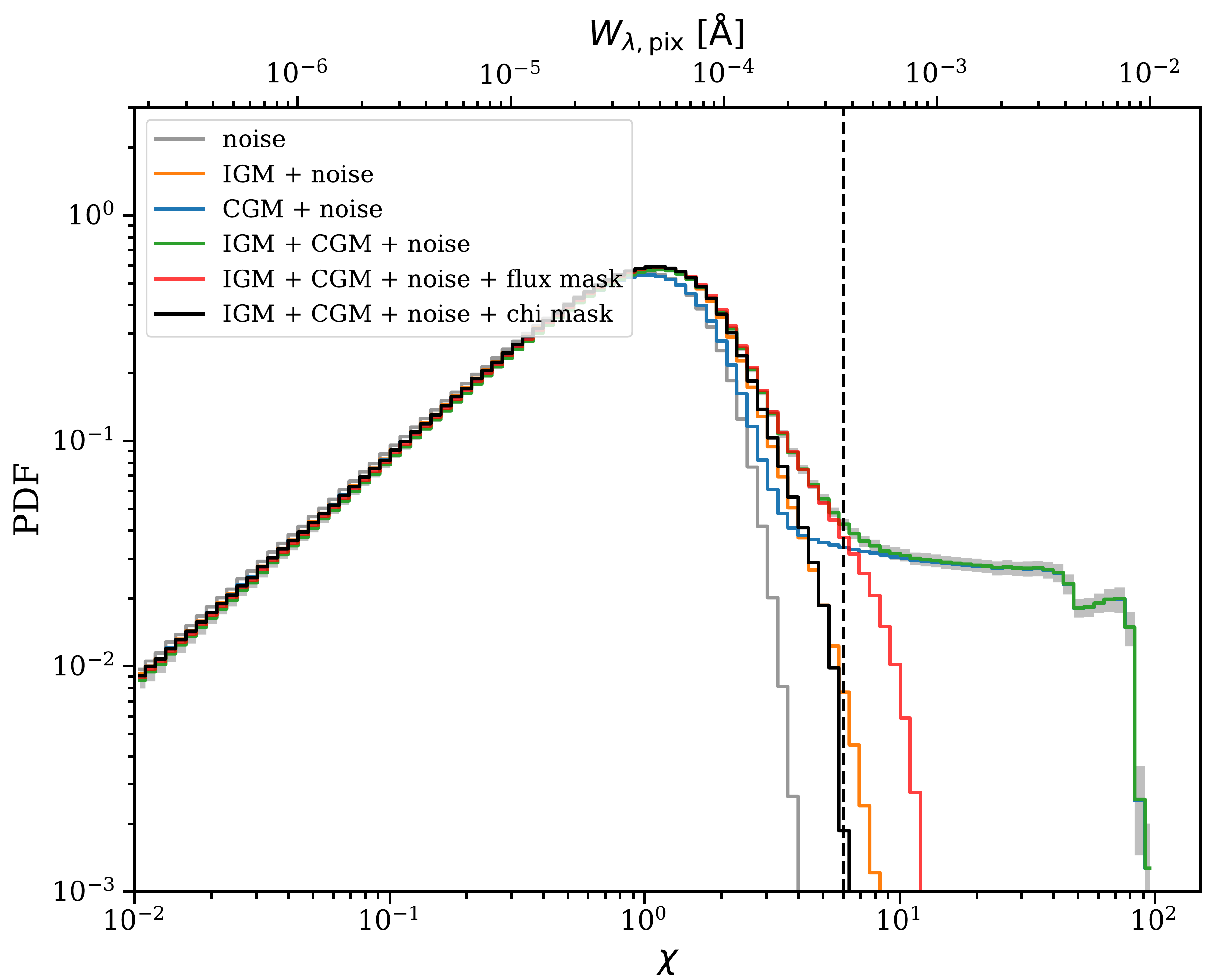}
\caption{Filtering schemes to remove CGM absorbers based on the flux decrement (left) and significance (right) PDFs. A random noise of SNR/pix = 50 has been added to all PDFs in both panels. The black PDFs show the resultant PDFs for the combined IGM + CGM + noise PDF after applying the cuts indicated by the vertical dashed lines, i.e. we remove all pixels with $1-F > 0.07$ in the left panel and all pixels with $\chi > 6$ in the right panel. These cuts remove large fluctuations caused by CGM absorbers, which are manifested in the tail of the IGM + CGM + noise PDF (green), in contrast to the IGM + noise PDF (orange) that drops off. Although the flux cut (red; right panel) removes a vast majority of the large fluctuations in the $\chi$ field, significant contamination remains when compared with the IGM + noise PDF (orange). The gray shaded regions on the IGM + CGM + noise PDF (green) are 1$\sigma$ errors of our mock dataset described in \S \ref{inference}. The IGM model here is the same as the inhomogenous IGM model in Figure \ref{pdf_wrange}.}
\label{cgm-masking}
\end{figure*}

The flux PDF provides a way to separate the CGM from the IGM absorbers because the flux PDF of the two populations have different shapes. The simplest cut involves a flux cut that filters out large $1-F$ values arising from the CGM, as shown by the left panel in Figure \ref{cgm-masking}. Here we have added noise with SNR/pix = 50 to our skewers. 

%% SST done, defined in first paragraph in 3.2.2
%% JFH YOu never defined W_lambda, pix

\begin{figure}
\centering
\includegraphics[width=\columnwidth]{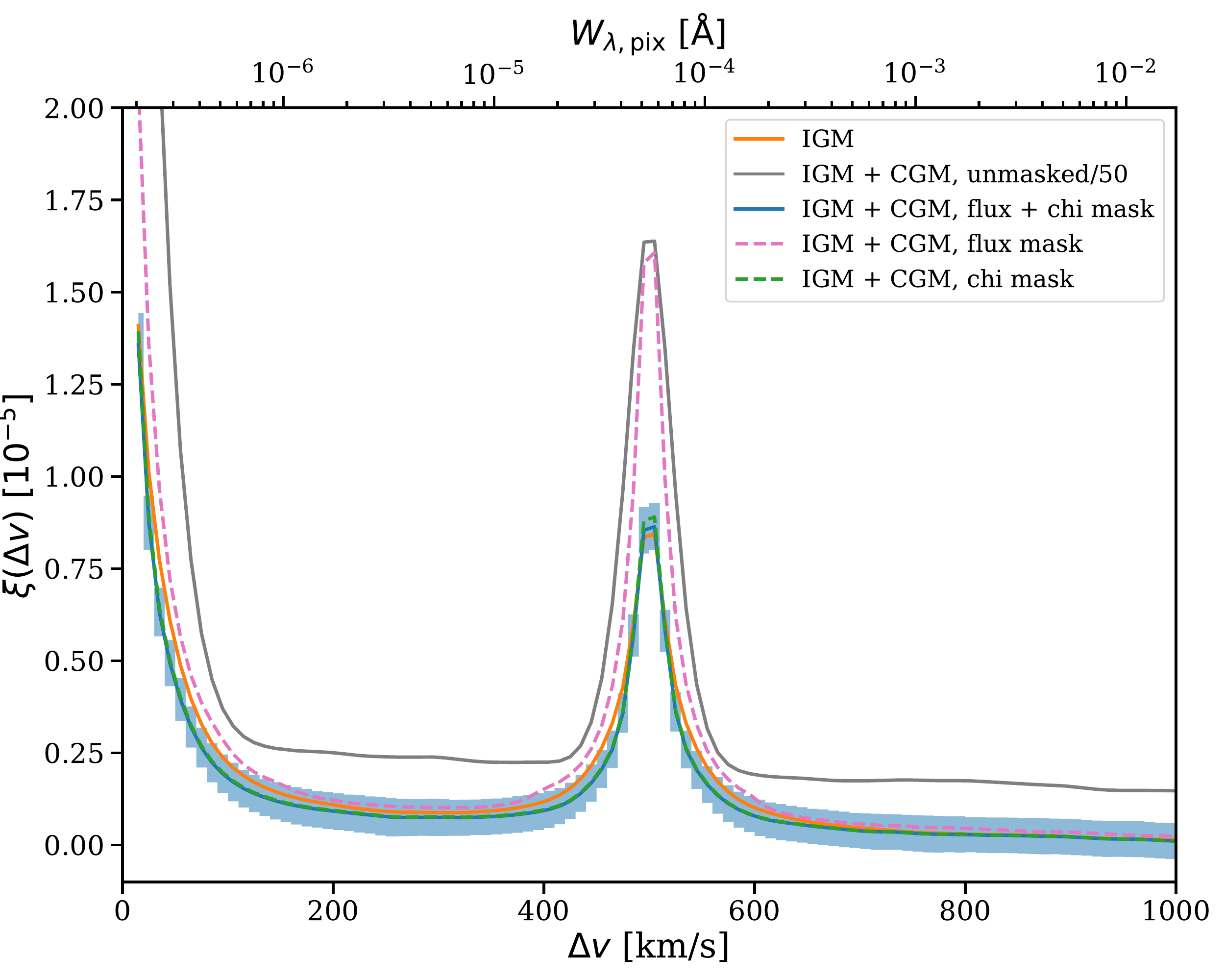}
\caption{Comparison of the \ion{C}{IV} forest correlation function before and after masking. Note that the unmasked IGM + CGM (gray) has been rescaled down by a factor of 50 to aid visualization. The effectiveness of the masking procedures is judged against the true signal from the pure IGM (orange), where we use the model with (log$\,M_{\rm{min}}$, $R$, [C/H]) = (9.50 $M_{\odot}$, 0.80 Mpc, $-3.50$). The cuts used in the masking are the same as in Figure \ref{cgm-masking}. Despite removing majority of the CGM signal, flux masking alone (pink dashed) results in residual bias in the masked spectra. Using either a chi cut only (green dashed) or combining the flux cut with a chi cut (blue) suppresses this bias and recovers the underlying IGM correlation function within the 1$\sigma$ errors of our mock dataset.}
\label{cf_mask}
\end{figure}

Although the noisy IGM PDF is indistinguishable from the noisy CGM PDF at small $1-F$ values, values with $1-F > 0.07$ in the combined noisy PDF mostly lie outside the IGM PDF and are due to CGM absorbers. As such, we can simply mask out pixels with $1-F > 0.07$, which only minimally reduces the number of IGM pixels in our mock spectra by 3\%. Figure \ref{cf_mask} shows the correlation function after masking out pixels with the flux cut. 
%% SST done
%% JFH2 This sentence would be easier to read if you stated linestyle and color that hte reader should focus on. That figure
%% has lots of curves. 
Note that the correlation function of the unmasked IGM+CGM (gray solid line) has been rescaled. The correlation function of the flux-masked pixels (pink dashed line) is biased slightly high compared to that of the pure IGM (orange solid line). Rather than arising from weak absorbers, which overlap with the IGM absorbers but are less abundant, this bias is instead due to unmasked pixels from the wings of the strong CGM absorbers. The flux cut manages to filter out the absorption cores of these absorbers but misses their shallow wings. To reduce this bias we can supplement the flux cut with another filtering method.
%To reduce this bias, we can use a stricter flux cut, e.g. filtering out pixels with $1-F > 0.04$ reproduces the IGM correlation function, but requires good understanding of the underlying noise, which is challenging to do with real data. Instead, we can supplement the flux cut with another filtering method. 

%% sst, done, modified the sentence.
%% JFH I don't understand why you are saying you need to understand the noise to make this cut. I don't agree with that. YOu simply plot the data and 
%% you can make this cut. I think your argument is instead that you would not be sure whether you are masking noise or signal, but I don't think that is the core
%% issue here. You can also mask your model in the same way that you mask your data. It is not clear to me that precise knowledge of the noise is really required. 
%% A more compelling arguement to pursue an additional cut is that you can do it trivially, i.e. you need to mask the wings of absorbers and the best way to do that
%% is identify absorbers via chi-filtering and then mask them. 

The second filtering scheme involves explicitly identifying CGM absorbers. In standard practice, this is done by convolving mock spectra with a matched filter $W(v)$ that corresponds to the transmission profile of a doublet \citep{ZhuMenard2013}, such that extrema in the significance field (\citetalias{Hennawi2020}),
%% SST done
%% JFH2 You've been referring to this as H2020, so be consistent in the referencing. 
%% SST done
%% JFH Maybe cite Hennawi 2020 here somewhere since it looks like you are attributing eqn. 13 to Zhu and Menard, which is not where it comes from. They 
%% convolve with a matched filter, but the definition of chi is from H2020
\begin{align}
\chi(v) = \frac{\int [1-F(v')]W(|v - v'|) dv'}{\sqrt{\sigma_F^2(v')W^2(|v - v'|) dv'}},
\end{align}
\noindent are assumed to be from an absorber. In the above definition, $\sigma_F^2$ is the flux variance due to noise. The matched filter $W(v)$ is $1 - e^{-\tau(v)}$, where $\tau(v)$ is the optical depth of a \ion{C}{IV} doublet with a Gaussian velocity distribution, assuming $N_\mathrm{\ion{C}{IV}} = 10^{13.5}$ cm$^{-2}$ and Doppler parameter 
%% SST done
%% JFH Make it more clear here that FWHM is the resolution of your spectrum and state what forward model you assumed. 
$b = \sqrt{2} \sigma = 6.02$ km/s assuming $\sigma = \mathrm{FWHM}/2.35$ and FWHM = 10 km/s for the resolution of our mock spectra in \S \ref{inference}.
%% SST looking at the COG, we're still technically at the linear part with b=6 km/s, though very close to the flat part. Remind me to send a plot of the filter profile (and COG plot), but the 1-F is still small (~ 0.12), so I assume we're fine? 
%% JFH Have you looked at this profile for W(v). I think the goal here is to guarantee that you are on the linear COG since those are the hardest absorbers to 
%% identify. I fear that given your parmeters (the small b-value), you may actually be on the flat part of the COG already. 
The right panel of Figure \ref{cgm-masking} shows the $\chi$-PDF of our mock spectra. Similar to the flux PDF, CGM absorbers give rise to large $\chi$ values. We can therefore remove them by (i) implementing a $\chi$ cut with $\chi > 6$ and (ii) removing pixels $\pm 200$ km/s around each extrema --- this is what we refer to as a ``chi cut''. Our final mask is a ``flux + chi'' mask, where ``flux + chi'' $\equiv$ flux cut OR chi cut. As shown in Figure \ref{cf_mask}, the correlation functions using the various masks (blue solid line) very closely match that of the pure IGM (orange solid line) within the measurement errors of our mock dataset in \S \ref{inference}. A flux, chi, and flux + chi mask removes 3\%, 35\%, and 35\% of the pixels in our spectra, respectively.
%% SST corrected
%% JFH2 I'm confused by these numbers. Is this the fraction retained or removed. You said earlier flux cut removes
%% 3% of pixels, but it cannot be that the chi-cut kills 65%. Also doesn't the match between the 35% imply that the 
%% chi-cut gets all the flux cut pixels. That is fine, I'm just checking. 

%% SST done, also explained chi cut vs flux + chi cut more explicitly. 
%% JFH I think it is more informative to state the number of pixels removed by flux masking and the number of pixels removed by flux masking OR chi (+ extrema masking).  
%% SST done
%% JFH Since you didn't explain how you compute measurement errors yet, I would forward cite to the section that does. 
Although masking reduces the overall pathlength and increases the statistical error, 
%% SST done
%% JFH statistical error
it does not bias correlation function measurements\footnote{This is only true if the CGM absorbers are randomly distributed, which we assume here. In reality, CGM absorbers preferentially reside in overdense regions, so masking them could slightly bias the correlation function. 
%% SST done
%% JFH2 Add one more sentence here making it clear that as long as one matches observations and simulations in the same
%% way the data-model comparison would still be well-posed and return unbiased parameter inference. 
However, as long as one applies the masking procedures consistently in simulations, the data-model comparison would still be valid and return unbiased parameter from the correlation function.}. Indeed, compared to the power spectrum, which in the simplest implementation would be sensitive to masking, the correlation function does not care about gaps in the spectra created from masking.
%% SST done
%% JFH Maybe add a footnote explaining that this is only true if the CGM absorbers are not strongly clustered (i.e. randomly distributed). Otherwise, you are 
%% techinically slightly biasing the results by masking CGM absorbers which could preferentially reside in overdense regions. 

For the particular IGM model we have chosen, the flux cutoff at $1-F > 0.07$ and the $\chi$ cutoff at $\chi > 6$ work well in removing the CGM absorbers and reproducing the pure-IGM correlation function. In real data, the choice of where to place the cutoffs can be made by computing the corresponding flux and $\chi$-PDFs and inspecting the region where the PDF starts to drop and then flatten. The choices of the cutoff values can also be forward-modeled in mock spectra and accounted for. As we have shown that CGM absorbers can be filtered out for our mock dataset with SNR/pix of 50, we did not include them in the next section where we perform statistical inference and evaluate the precision of our model constraints from mock observations. In Appendix A, we investigate how lower SNR data can potentially affect the masking procedures. 

Lastly, as Figure \ref{pdf_mrz} shows, changes in the IGM model parameters noticeably changes the flux PDF, which suggests that one can potentially constrain the model parameters using the flux PDF. In practice however, this also requires knowing the noise properties of the underlying data, since the final PDF is affected by the shape of the noise PDF, as Figure \ref{cgm-masking} shows.

\subsection{Inference and constraints}
\label{inference}
In this section, we estimate the precision of constraints that can be achieved for a realistic mock dataset. We ignore the effects of CGM absorbers, as they can be filtered out relatively easily, as we have shown in \S \ref{maskingcgm}. 

For our mock dataset, we consider $n_{\mathrm{QSO}}$ = 20 quasar spectra that are convolved with a spectral resolution of a Gaussian line spread function with FWHM = 10 km/s ($R$=30,000; achievable with Keck/HIRES or VLT/UVES). Gaussian random noise with $\sigma = $ (SNR)$^{-1}$ and SNR = 50 are added to each pixel. Our spectral sampling is 3 pixels per resolution element of width the FWHM  specified above.  
%% SST done
%% JFH Wherever you mention Keck/HIRES you should also probably also mention VLT/UVES. We may be writing proposals to use both instruments. 
The redshift extent from the \ion{C}{IV} to the Ly$\alpha$ emission lines is $dz = 1.18$. As our goal is to evaluate how well correlation function analysis can constrain IGM metallicity and enrichment topology, we choose a large \ion{C}{IV} forest pathlength of $dz=1.0$, which corresponds to 93 skewers from our simulation for a desired total pathlength of $\Delta z$ = 20. A smaller $dz$ is more appropriate for studies of redshift evolution, which is not the focus of this paper. 

We create $10^6$ realizations of our dataset for each model in our (log$\,M_{\mathrm{min}} \times R \times$ [C/H]) = (26 $\times$ 30 $\times$ 26) 3D grid, by randomly drawing 93 skewers each time without replacement, and compute their correlation functions and covariance matrices.
%% SST done
%% JFH covariances --> covariance matrices
We perform Markov Chain Monte Carlo (MCMC) assuming a multivariate Gaussian likelihood
%\begin{align}
\begin{multline}
L(\hat{\xi}(dv) | \mathrm{log}M_{\mathrm{min}}\mathrm{, R, [C/H]}) =\\ \frac{1}{\sqrt{(2\pi)^k \mathrm{det}(\mathrm{\mathbf{C}})}}\mathrm{exp}(-\frac{1}{2}\mathbf{d}^T \mathbf{C^{-1}} \mathbf{d}), 
\end{multline}
%\end{align}
\noindent where \textbf{d} = $\hat{\xi}(dv) - \xi(dv | \mathrm{log}M_{\mathrm{min}}\mathrm{, R, [C/H]})$ is the difference between the measured correlation function $\hat{\xi}(dv)$ and the model correlation function $\xi(dv |\mathrm{log}M_{\mathrm{min}}\mathrm{, R, [C/H]})$. The measured correlation function is averaged over our mock dataset of 93 skewers and the model correlation function is averaged over the total 10,000 skewers. The velocity bin $dv$ is set to be equal to the FWHM, $dv = 10$ km/s, and the correlation function is measured over $v_\mathrm{min} = 10$ km/s to $v_\mathrm{max} = 2000$ km/s. We compute the covariance matrix element as 
\begin{align}
    C_{ij} %&= \langle d_i d_j\rangle \\
    &= \langle [\hat{\xi}(dv) - \xi(dv)]_i [\hat{\xi}(dv) - \xi(dv)]_j \rangle,
\end{align}
\noindent where $i$ and $j$ indicate different bins of $dv$ and the angle brackets mean averaging over $10^6$ realizations of each mock dataset.

We use \texttt{EMCEE} \citep{Foreman-Mackey2013} to perform our MCMC analyses. We assume flat log priors for log$\,M_{\rm{min}}$ and [C/H] and a flat linear prior for $R$ extending over the entire range of the grid, i.e. from 0.8 $M_{\odot}$ to 11.0 $M_{\odot}$ for log$\,M_{\rm{min}}$, from $-4.5$ to $-2.0$ for [C/H], and from 0.1 to 3.0 Mpc for $R$. As our model grid is coarse, to speed up the MCMC sampling process, we interpolate the likelihood computed at our initial grid onto a finer grid, for which we use the \texttt{ARBInterp}\footnote{\url{https://github.com/DurhamDecLab/ARBInterp}} tricubic spline interpolation\footnote{While most tricubic interpolations split the problem into three one-dimensional problems, this method is intrinsically three-dimensional \citep{LekienMarsden2005}. We initially experimented with 3D linear interpolation using Scipy \texttt{RegularGridInterpolator}, but found that the interpolation is not smooth and results in the MCMC being sensitive to the interpolation.} routine \citep{Walker2019}.
%% SST done, small changes, added "as our model grid is coarse, to speed up..."
%% JFH The finer grid is just a trick to make the evaluation fast. The main point is that your model grid is coarse whereas you need to be able to evaluate anywhere
%% in param space to perform the MCMC. 

In principle one can also infer the effective metallicity given the inferred input metallicity and morphological parameters. For this, we first evaluate the effective metallicity at each model grid point and create a lookup table, i.e. [C/H]$_{\rm{eff}}$(log$M_{i}$, $R_{j}$, [C/H]$_{k}$). We then interpolate  
%% SST done
%% JFH You don't interpolate onto the lookup table, you interpolate from the lookup table. "We then interpolate from this look up table onto the MCMC chain output locations"
from this lookup table onto the MCMC chain output to derive the posterior PDF of the effective metallicity and the resultant uncertainty. In the results below, note that the corner plots for the effective metallicity are not direct outputs from the MCMC, but instead derived from them. 

\begin{figure*}
\includegraphics[height=0.27\textheight, width=0.6\textwidth]{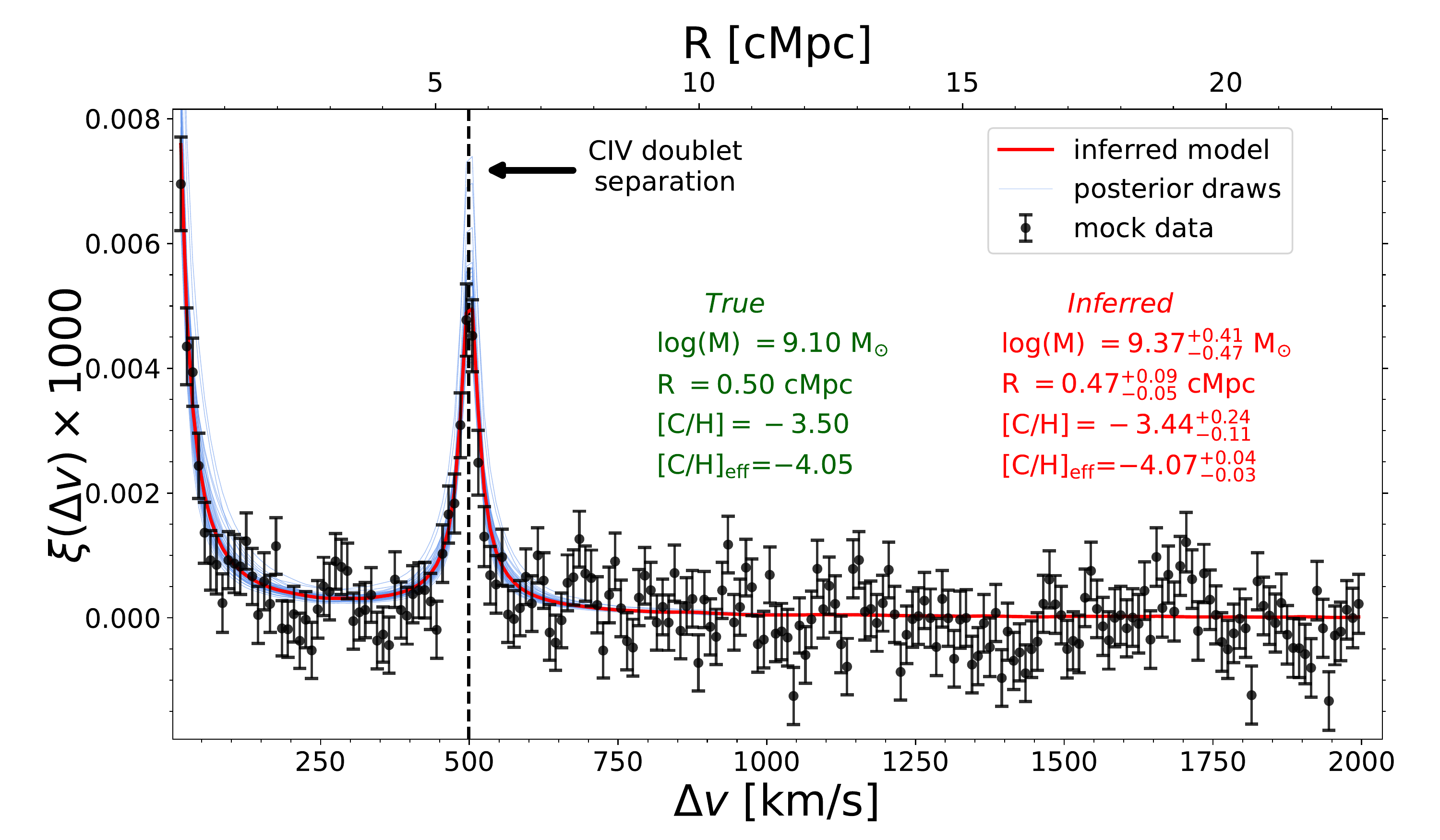}
\includegraphics[height=0.27\textheight, width=0.39\textwidth]{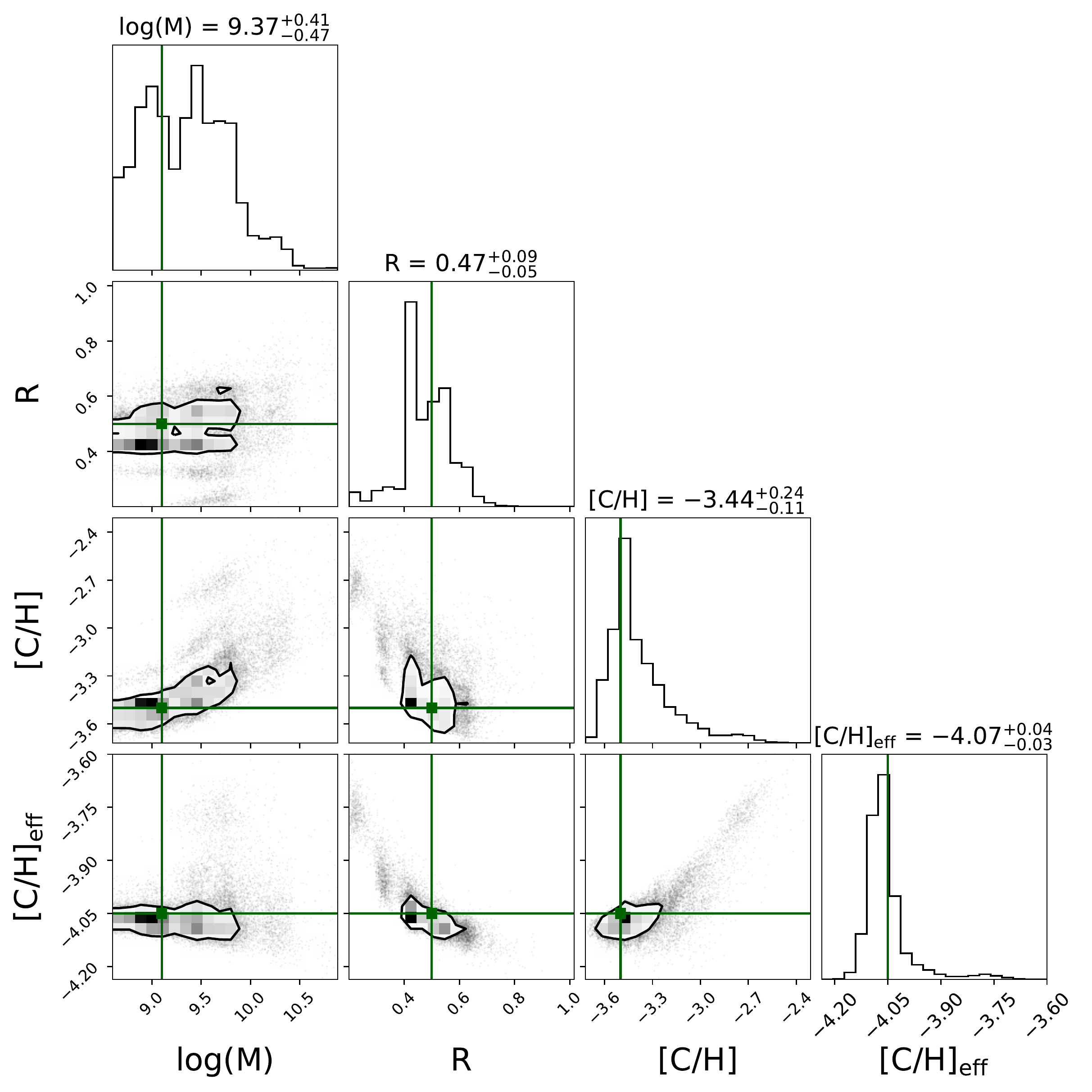}
\caption{Correlation function measured from our mock dataset (left) and the inference results (right). The mock dataset is characterized by 20 QSO spectra, each with a \ion{C}{IV} pathlength of $dz=1.0$, FHWM = 10 km/s, and SNR/pix = 50. Also plotted in the correlation function plot are fifty random draws (thin blue lines) and the mean inferred model (red line and red text) from the MCMC posterior distribution. The effective metallicity [C/H]$_{\mathrm{eff}}$ for the true model is calculated according to Eqn (3). The right panel shows the corner plot from MCMC sampling, where the true models are indicated by the green markers and lines. Note that the corner plot for [C/H]$_{\mathrm{eff}}$ is indirectly derived from the MCMC outputs of the other parameters (see text). The inferred parameters shown on the top of the corner plots represent the median and the 68\% credible intervals. For this IGM model, we expect to simultaneously measure the values of log$\,M_{\rm{min}}$ to a precision of $\sim 0.44$ dex, [C/H] to a precision of $\sim 0.2$ dex, and $R$ to within $\sim 15\%$. The effective metallicity can be constrained indirectly within less than 0.05 dex.}
\label{mcmc1}\
\end{figure*}

\begin{figure*}
\includegraphics[height=0.27\textheight, width=0.6\textwidth]{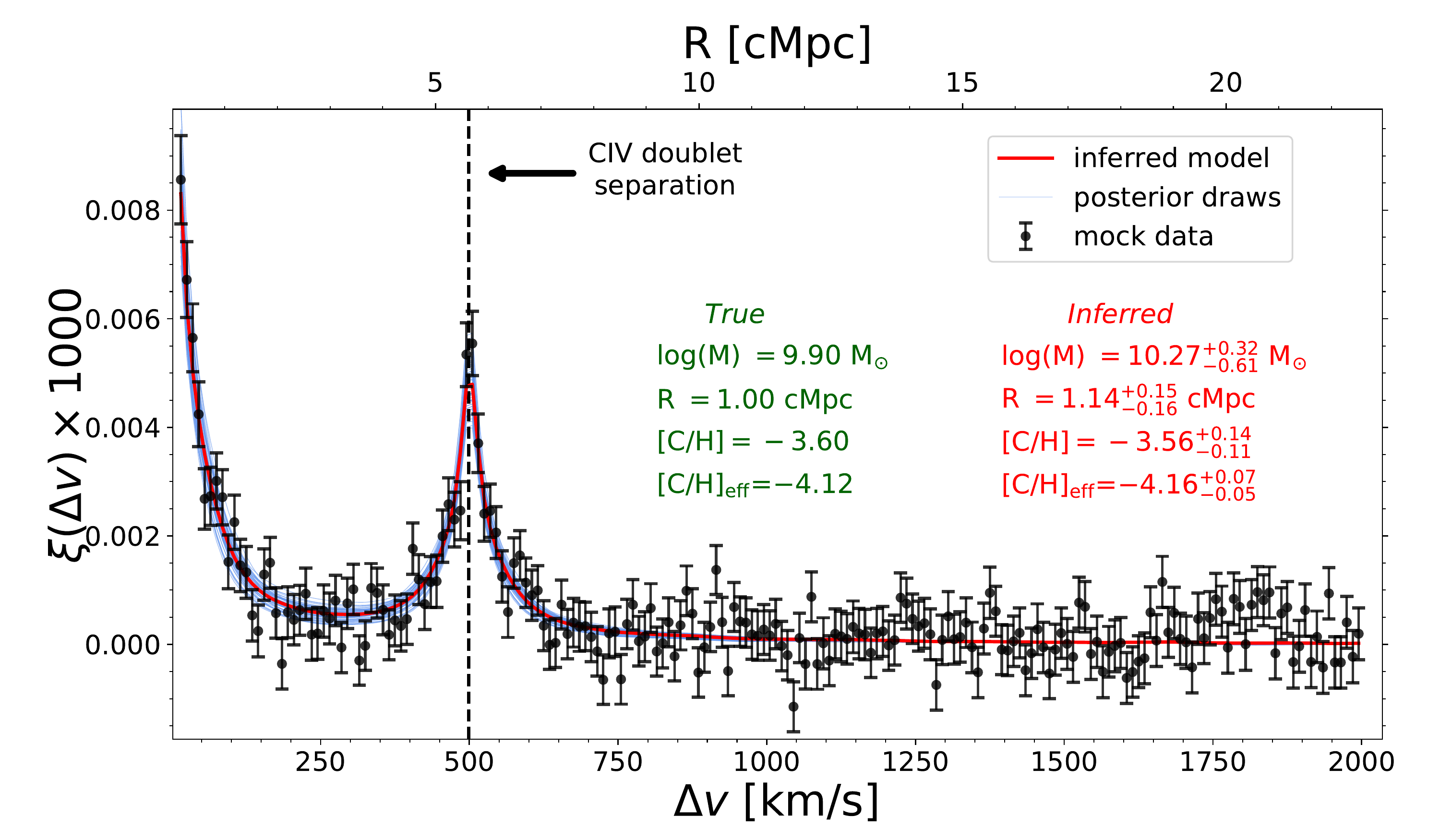}
\includegraphics[height=0.27\textheight, width=0.39\textwidth]{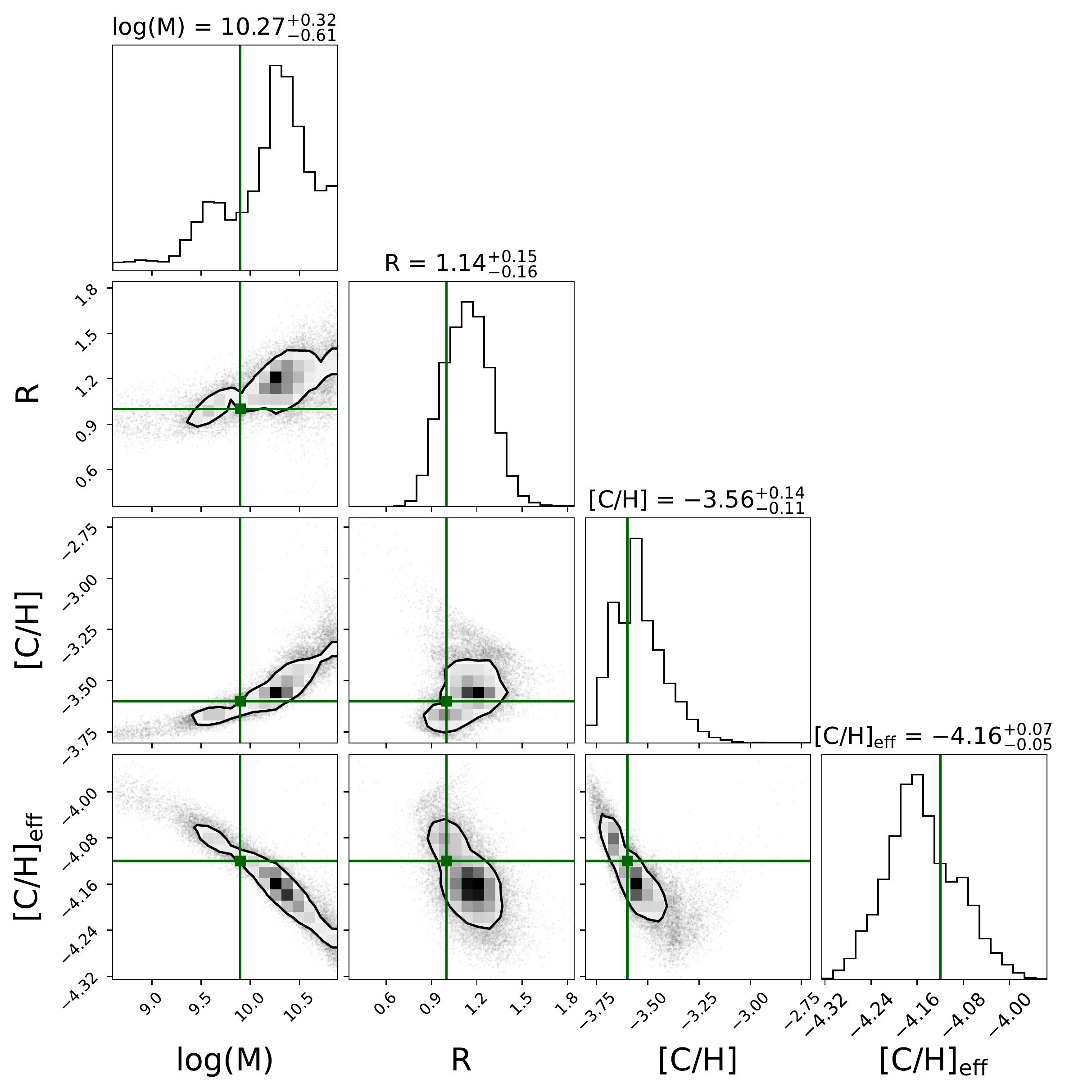}
\caption{Same as Figure \ref{mcmc1} but for a different IGM model with larger log$\,M$ and $R$ and smaller [C/H]. We find comparable precisions as Figure \ref{mcmc1}, where assuming our mock dataset, log$\,M_{\rm{min}}$ is measured to within $\sim 0.45$ dex, [C/H] to within 0.125 dex, and $R$ to within 13.5\% at 1$\sigma$. The effective metallicity is indirectly constrained to less than 0.1 dex. }
\label{mcmc2}
\end{figure*}

Figure \ref{mcmc1} shows the correlation function measured from the mock dataset and the resulting parameter constraints for a model with (log$\,M_{\rm{min}}$, $R$, [C/H]) = (9.10 $M_{\odot}$, 0.50 cMpc, $-3.50$), where the corresponding ($f_V$, $f_m$, [C/H]$_{\mathrm{eff}}$) = (0.066, 0.28, $-$4.05).
%% SST done
%% JFH Quote the corresponding f_v, f_M and [C/H]_eff for these models in the text. 
Given a real dataset resembling our mock version, the values of log$\,M_{\rm{min}}$ and metallicity can be measured to a precision of $\sim 0.44$ and $\sim 0.2$ dex, respectively, while $R$ is expected to be constrained to within $\sim 15\%$. Figure \ref{mcmc2} shows the results for a different model, (log$\,M_{\rm{min}}$, $R$, [C/H]) = (9.90 $M_{\odot}$, 1.00 cMpc, $-3.60$) with the corresponding ($f_V$, $f_m$, [C/H]$_{\mathrm{eff}}$) = (0.11, 0.31, $-$4.12). Here, log$\,M_{\rm{min}}$ is measured to within $\sim 0.45$ dex, [C/H] to within 0.125 dex, and $R$ to within 13.5\%. That the metallicity can be constrained more precisely is likely due to it being less degenerate with the other parameters at these parameter locations. The morphological parameters have larger errors due to their more complex effects on the amplitude and shape of the correlation function, compared to the quadratic amplitude scaling with metallicity (see Figure \ref{cf_mr_logZ}).
%% SST done
%% JFH2 Maybe put in a reference to Figure 6 after this statement. 
%Since the correlation function has a simple dependence on the metallicity, its value is able to be measured the most precisely. 
%% SST Not sure how to better phrase it; see attempt above "That the metallicity ..."
%% JFH I don't think a simple dedpendence is the point. The point is I think how strong the dependencies are and whether there are strong degeneracies with other
%% parameters. Given the quadratic amplitude scaling with the metallicity parameter that is probably what you are seeing here, but why this is not strongly
%% degenerate with changing the f_M is not so obvious to me. It is probably a statement about the range of parameters considered and their dependencies, i.e. 
%% if the amplitude changes by a factor of X when you chnage logZ, how much would you need to change logM and R to get the same factor of X amplitude increase. 
In both models, the effective metallicities are (indirectly but precisely) constrained to within less than 0.1 dex.

Finally, it is worth asking what upper limit we can set on the IGM metallicity in case of a null detection of the correlation function. 
%% SST done (need to double-check)
%% JFH For quoting the upper limit, what if we simply restrict to a uniform IGM. Then the analysis is simple and analogous to what I did in my paper. I think 
%% you can mock this up by simply restricting to models with f_M = 1.0
We estimate the upper limit by restricting to models with a uniformly-enriched IGM. In this case, the inference problem now becomes one-dimensional with [C/H] being the only parameter. Assuming a flat prior and randomly choosing a mock dataset among our 10$^6$ realizations, we compute the posterior numerically for a model with [C/H] = $-5.0$ and obtain an upper limit of [C/H] $< -4.35$ at 95\% confidence. 

\section{Discussions and conclusions}
\label{conclusion}
We investigate the correlation function of the \ion{C}{IV} forest as a probe of IGM metallicity and enrichment topology. We generate models of inhomogeneous enrichment using the halo catalogs from the $z=4.5$ snapshot of the \texttt{Nyx} hydrodynamic simulation, whereby halos above a minimum mass log$\,M_{\rm{min}}$ enrich their environments to a constant metallicity [C/H] out to a maximum radius $R$. We simulate the $z=4.5$ \ion{C}{IV} forest with the \texttt{Nyx} simulation and compute the ionization fraction of \ion{C}{IV} using \texttt{CLOUDY}. Skewers of the \ion{C}{IV} forest are computed for each inhomogeneous enrichment model by interpolating \texttt{CLOUDY} ionization fraction outputs based on \texttt{Nyx} density and temperature fields. 

The two-point correlation function (2PCF) of the \ion{C}{IV} forest skewers has a clear peak at $dv = 498$ km/s due to the doublet separation of the \ion{C}{IV} absorption line. The amplitude of this peak increases quadratically with metallicity. 
%% SST corrected
%% JFH I think the optical depth is linear, but the correlation function is quadratic. 
The enrichment morphology parameters affect both the shape and amplitude of the 2PCF, where increasing log$\,M_{\rm{min}}$ and $R$ individually leads to an increase in power at both large and small scales. Since different combinations of log$\,M_{\rm{min}}$ and $R$ give rise to different values of mass- and volume-filling factors, the effect of enrichment morphology can also framed in terms of the filling factors.

Measurements of the IGM enrichment topology remain sparse and poorly-constrained. \cite{Booth2012} concluded that $f_V > 0.1$ and $f_m > 0.5$ are required to match the observed median \ion{C}{IV} optical depth as a function of $\tau_{\mathrm{HI}}$ at $z=3$, which are satisfied when the IGM is predominantly enriched by low-mass ($< 10^{10} M_{\odot}$) galaxies out to $r \geq 100 $ proper kpc. The importance of low-mass galaxies for enriching the IGM is also found by other theoretical studies (e.g. \citealp{Samui2008,Oppenheimer2009,Wiersma2010}). On the other hand, by computing the clustering of discrete absorbers at the same redshift ($z=3$), \cite{Scannapieco2006} instead found that high-mass ($\approx 10^{12} M_{\odot}$) galaxies are mainly responsible for IGM enrichment out to $r \geq 500 $ proper kpc. We estimate the precision to which the IGM metallicity and enrichment topology can be inferred using mock observations of 20 quasar spectra, where each spectrum is convolved to a resolution of FHWM = 10 km/s (equivalent to $R$=30,000 of Keck/HIRES or VLT/UVES) and has a SNR/pix = 50, for a total pathlength of $\Delta z$ = 20. The two IGM models shown in Figure \ref{mcmc1} and Figure \ref{mcmc2} are consistent with the best-fit model of \cite{Booth2012}, although note that their work is at a much lower redshift than ours. We find that we can constrain the metallicity to a precision of $\sim 0.2$ dex at 1$\sigma$, while log$\,M_{\rm{min}}$ and $R$ can be constrained to within $\sim 0.4$ dex and $\sim 15\%$, respectively. Not only is our method different than those employed in existing studies, it is currently the state-of-the-art in simultaneously constraining the IGM metallicity and enrichment topology to high precision\footnote{Our mock dataset is higher in both number and data quality than \cite{Booth2012} but comparable to \cite{Scannapieco2006}.}\footnote{By incorporating MCMC inferencing and letting metallicity be a free parameter in addition to topology, our work is more rigorous and expansive than  \citealp{Booth2012} and \citealp{Scannapieco2006}, who first fixed metallicity before deriving limits/estimates on the enricment topology via simple model fitting to data.}. We plan to apply our correlation function method to a real dataset in future work in the hope of potentially alleviating this disagreement. 

We discuss how CGM absorbers near galaxies can bias clustering measurements of IGM absorbers, and investigate their effects by modeling their abundance based on observational constraints and injecting them into our mock spectra. We propose methods based on the flux PDF that can effectively remove CGM absorbers. While the flux distribution of IGM absorbers peaks at small $1-F$ values of $0.003 - 0.004$ and exponentially drops off at large values, the flux distribution of CGM absorbers remains mostly flat within the overlapping region. Depending on the IGM model, the IGM absorbers dominate over CGM absorbers by $\sim 20\%$ to a factor of a few tens at the peak. For a plausible IGM model with (log$\,M_{\rm{min}}$, $R$, [C/H]) = (9.50 $M_{\odot}$, 0.80 cMpc, $-3.50$) corresponding to ($f_V$, $f_m$, [C/H]$_{\mathrm{eff}}$) = (0.12, 0.34, $-3.97$), 
%% SST done
%% JFH2 Can you quote effective metalllicity for this model as well
IGM absorbers are $\sim 7-8$ times more abundant. While IGM absorbers dominate even more in models with higher filling fractions, up to $\sim 80$ times for a uniformly-enriched IGM with [C/H] = $-3.50$, 
%% SST done
%% JFH2 I'm not sure what a uniform IGM refers to here? Do yuu mean uniformly enriched and if so at C/H???
CGM absorbers start to become comparable with IGM absorbers for models with ($f_V$, $f_m$) = (0.001, 0.05) and lower.   
%% SST added more details above
%% JFH I suggest to rephrase this sentence above. Maybe quote a fiducial number for a plausible model, then 
%% state that this however depends on the model parameters (and maybe state the range). You are basically saying
%% the IGM dominates by a  factor of 1.2 to 30, which seems like a not very informative statement, i.e. 
%% better to state the expectation and then the range. 
Due to the distinct shapes of the IGM vs. CGM flux PDFs, we can remove CGM absorbers using a simple flux threshold. We also investigate an additional filtering scheme that automatically identifies CGM absorbers using their significance or $\chi$ field, whereby large $\chi$ values are attributed to CGM absorbers. By combining the flux threshold cutoff with a $\chi$ cutoff (including masking pixels around each extrema of the doublet), we can effectively mask out CGM absorbers and easily recover the underlying IGM correlation function from the masked spectra, since the correlation function is not affected by gaps in the spectra due to masking (as opposed to the power spectrum).

%% SST done
%% JFH Rather than gas density, I would focus on column density since it is an observable.

%% JFH2 In this paragraph you could also mention the CIV-Ly-a cross-correlation briefly and what we might learn from that. 
Another method to differentiate CGM from IGM absorbers is to perform a cut based on gas density. The gas density is related analytically to absorption in the Ly$\alpha$ forest, so Ly$\alpha$ forest absorption can be used as a proxy for density in real observations. In practice, since column density is an observable while density is not, it is easier to mask based on \ion{H}{I} column density, where high \ion{H}{I} column densities ($N_{\ion{H}{I}} > 10^{14}$ cm$^{-2}$) would likely correspond to CGM absorbers. One can also consider cross-correlating the \lya\ forest with the \ion{C}{IV} forest. In this work, we focus on the autocorrelation of the \ion{C}{IV} forest as a probe of the background IGM metallicity and ignore how metallicity varies as a function of gas density. Cross-correlating the \lya\ forest with the \ion{C}{IV} forest connects  the gas metallicity with the gas density and provides a more detailed picture of the IGM enrichment. This is motivationally similar to the pixel optical depth (POD) method that maps out the optical depth of metals as a function of the \ion{H}{I} optical depth. The difference is that while the POD method measures this relation at zero velocity lag, i.e. in the same gas that gives rise to both the \ion{H}{I} and metals, cross-correlation allows one to map out the relation over all velocity lags, thereby giving one more handle on the enrichment topology. 

%% SST all comments addressed above
%% JFH I think you can state high HI column densities would correspond to CGM absorbers
%For instance, one can remove all pixels that have a Ly$\alpha$ absorption corresponding to a gas overdensity of $\Delta=100$ or more. 
%% JFH again in practice this would be column density 

%Cutting based 
%% JFH I think you mean masking not cutting, and I don't know that we want to cut on flux, rather we would 
%% probably try to identify strong absorbers based on constraints from e.g. the Ly series. Recall that the 
%% Ly-a forest saturates at mean density, so cutting on flux is not a good idea. 
%on Ly$\alpha$ forest flux requires one to spectrally cover both the Ly$\alpha$ and the metal line forests,

%% JFH Having this joint coverage is not such a big deal, so maybe not worth mentioning. 
%in addition to cleaning up the Ly$\alpha$ forest from foreground contaminations. 

We did not consider foreground metal-line contamination in this work. Being one of the reddest metal lines and compared to bluer lines, \ion{C}{IV} does not suffer from significant foreground contamination. However, lower redshift redder lines such as \ion{Fe}{II} $\lambda1608$\r{A} and \ion{Mg}{II} $\lambda2796$\r{A} 
%% SST don't understand why the bluer Si II 1527 can contaminate the CIV forest. 
%% JFH You are right SiII 1527 cannot contaminate CIV, only redder lines. 
%% JFH See https://ui.adsabs.harvard.edu/abs/2018JCAP...05..029B/abstract 
%% for the relevant contaminants.
and $\lambda2804$\r{A} lines can enter the \ion{C}{IV} forest window and contaminate the resulting signal. Drawing inspiration from how Ly$\alpha$ forest cosmology handles metal line contamination (e.g., \citealp{McDonald2006,Palanque-Delabrouille2013}), one can first define wavelength windows that are redder than the \ion{C}{IV} line, measure the correlation function within these windows using lower-redshift quasars, and finally subtract the measured correlation function from the higher-redshift \ion{C}{IV} forest correlation function.
%% SST corrected above
%% JFH Actually this is not quite right. What one does is remove foreground contamination by subtracting
%% the correlation function from lower-redshift quasars from the signal from higher-z quasars. 
Conventional methods of identifying individual lines, e.g. exploiting the separation of a foreground metal doublet and searching for additional transitions at the same redshift of the first metal line \citep{Lidz2010}, can also be applied to individual spectra to remove foreground contamination. 
%% JFH You can also say that our PDF approach would naturally also mask these CGM lines, but not IGM lines.
%% Although I think they will be a lot weaker from the IGM. 

The auto-correlation of the metal-line forest presents a new avenue to constrain the enrichment history of the IGM, providing precise and simultaneous constraints on the IGM metallicity and enrichment topology. In a future work, we plan to apply this method to a sample of archival echelle quasar spectra.
%% SST done
%% JFH Also possibly UVES. I might just say archival echelle spectra of quasars. 
Viewed along current constraints obtained with existing methods such as the pixel optical depth method and standard Voigt profile fitting, this future measurement will shed light on the background IGM metallicity and cosmic enrichment history.
%% SST done
%% JFH2 "the Population III stars" things feels like a throw-away statement that can be misunderstood. Maybe say "shed light on the background IGM metallicity and 
%% hence the cosmic enrichment history" or something. I do think we could constrain PopIII enrichment possibly, but this would be hard, i.e. using relative
%% abundances and we certainly didn't demonstrate that here. 
There is also the possibility to extend the measurement to higher redshifts and investigate the evolution in the IGM metallicity. Higher-redshift measurements will be more challenging due to the decreasing metallicity and decreasing telescope sensitivity in the NIR (e.g., at $z=6$, the \ion{C}{IV} forest redshifts into the $Y$ band), so having a larger quasar sample and more sensitive space-based observations (e.g. with JWST) is ideal. Application of the method to other metal species provides complementary constraints on other aspects of IGM physics. For instance, clustering of the \ion{Mg}{II} forest allows one to constrain the neutral hydrogen fraction during reionization \citetalias{Hennawi2020}. Through measurements at different redshifts that probe the appearance of low-ions and the concurrent disappearance of high-ions with increasing redshift, it might be possible to observe the phase transition in the IGM due to reionization \citep{Oh2002} as well as to constrain the relative abundance of these ions (e.g. \citealp{Cooper2019,Becker2009}).
%% SST done
%% JFH2 Make it more clear in the last sentence that you would be looking for the low-ions to appear and the high-ions to dissappear as you probe to higher-z. 
%% JFH I would mention the idea of cross-correlating the CIV and HI forests and what we can learn from that. 
%% JFH I would mention the synergy with MgII forest, i.e. seeing the IGM transition from being ionized to neutral. You should also discuss the possibility of studying the CIV forest at higher-z than considered here
%% i.e. 4.5 < z < 6 to try to constrain metallicity at even higher-z. 

\section*{Acknowledgment}
We acknowledge helpful conversations with the ENIGMA group at UC Santa Barbara and Leiden University, and Frederick Davies at MPIA. 
%% SST done
%% JFH2 ENIGMA group is at UCSB and Leiden. If you want to thank MPIA people, I'd mention them. 
We especially thank Farhanul Hasan for sharing his data on CGM absorbers and Paul Walker for help with his tricubic spline interpolation code. 

This project has received funding from the European Research Council (ERC) under the European Union’s Horizon 2020 research and innovation program (grant agreement No 885301). JFH acknowledges support from the National Science Foundation under Grant No. 1816006. SEIB acknowledges funding from the European Research Council (ERC) under the European Union’s Horizon 2020 research and innovation programme (grant agreement No. 740246 ``Cosmic Gas''). Calculations presented in this paper used resources of the National Energy Research Scientific Computing Center (NERSC), which is supported by the Office of Science of the U.S. Department of Energy under Contract No. DE-AC02-05CH11231.

This research made use of \texttt{Astropy}\footnote{\url{http://www.astropy.org}}, a community-developed core Python package for Astronomy \citep{Astropy2013,Astropy2018}, \texttt{SciPy}\footnote{\url{https://scipy.org}} \citep{Scipy2020}, and the \texttt{ARBInterp}\footnote{\url{https://github.com/DurhamDecLab/ARBInterp}} tricublic spline interpolation routine.

\section*{Appendix A}
\begin{figure*}
\centering
\includegraphics[width=\columnwidth]{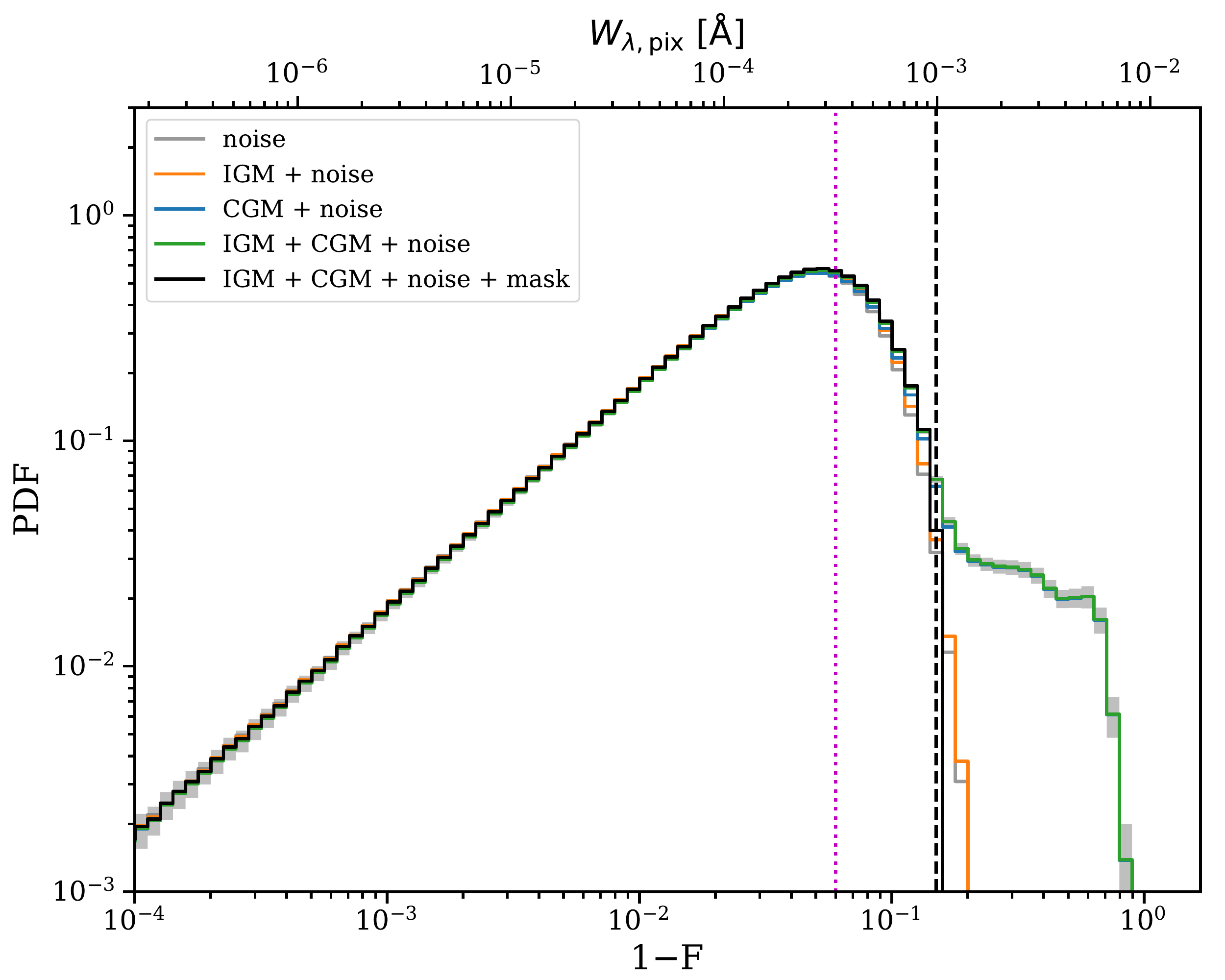}
\includegraphics[width=\columnwidth]{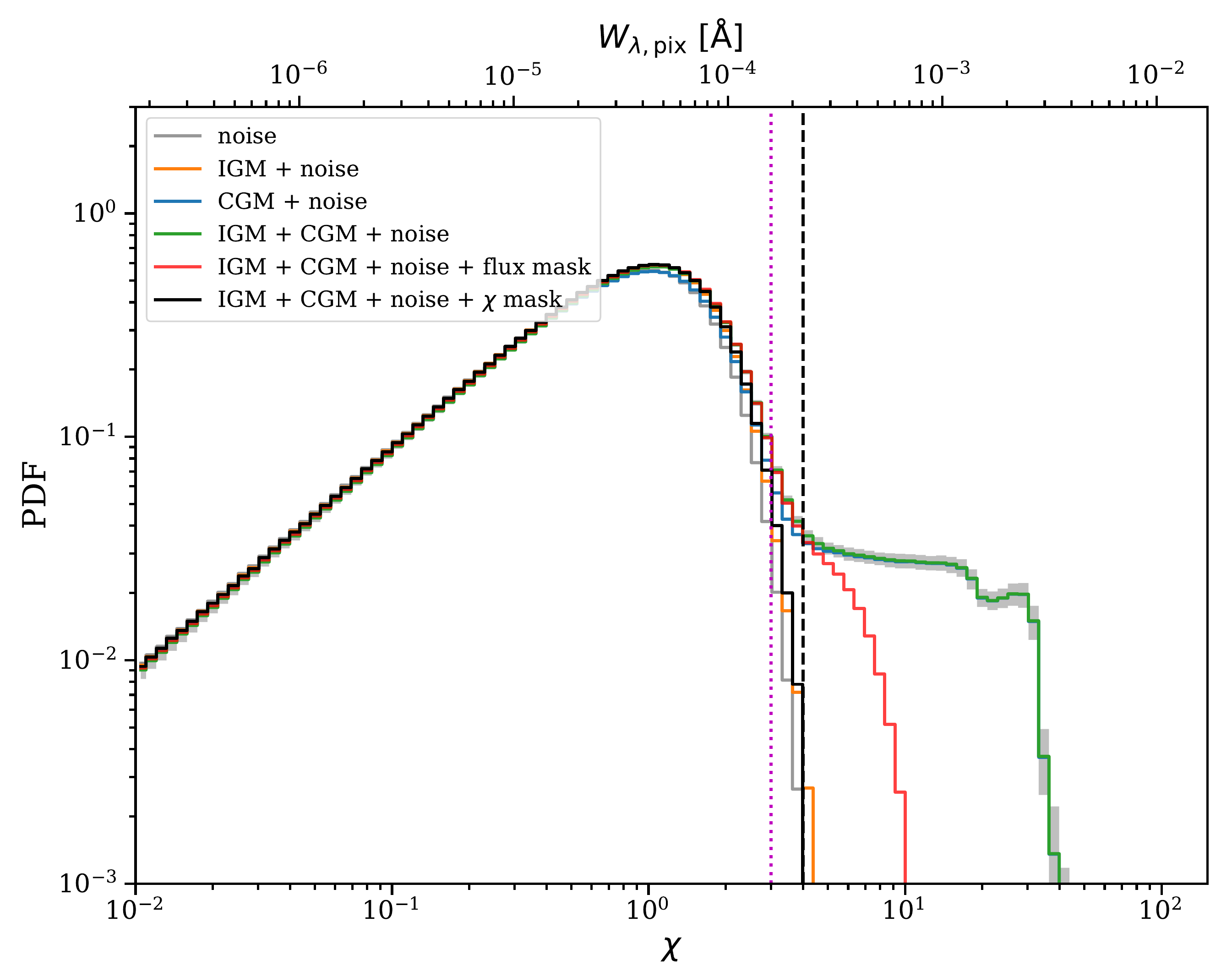}
\caption{Same as Figure \ref{cgm-masking} and using the same IGM model, but with added Gaussian random noise resulting in SNR/pix = 20 instead. The cutoff thresholds indicated by the vertical black lines, located at $1 - F = 0.15$ and $\chi = 4$, result in the masked correlation functions shown in Figure \ref{cf_mask_snr20}, while the more aggressive cutoffs indicated by the the vertical magenta lines, located at $1 - F = 0.06$ and $\chi = 3$, give rise to the correlation functions shown in Figure \ref{cf_mask_snr20_2}. }
\label{cgm-masking-snr20}
\end{figure*}

\begin{figure}
\centering
\includegraphics[width=\columnwidth]{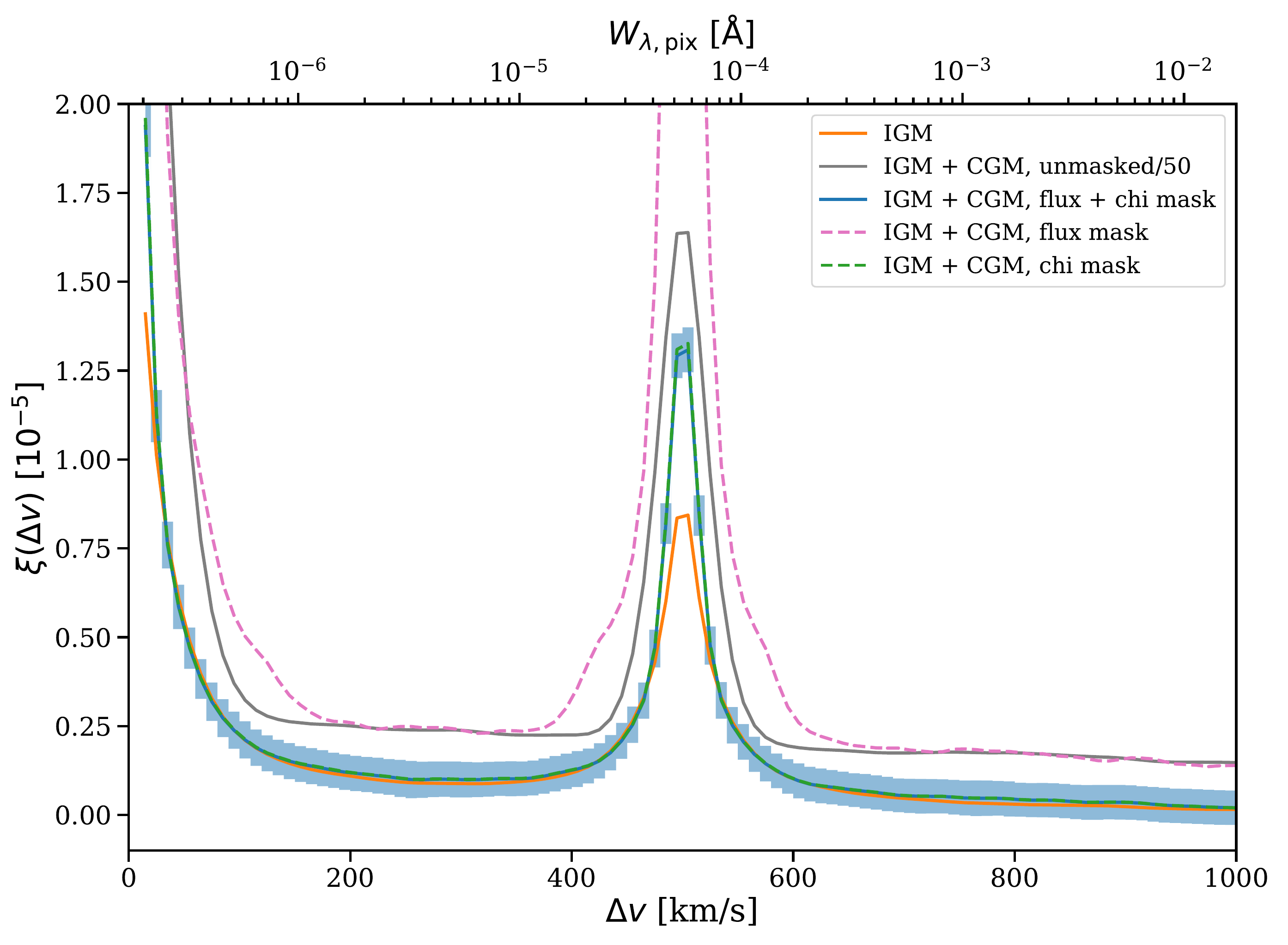}
\caption{Comparison of the \ion{C}{IV} forest correlation function before and after masking at the vertical black lines shown in Figure \ref{cgm-masking-snr20}. Note that the unmasked IGM + CGM (gray) has been rescaled down by a factor of 50 to aid visualization. Flux masking alone results in a signicantly biased correlation function (dotted line), and residual bias remains even after combining the flux cut with a significance cut.}
\label{cf_mask_snr20}
\end{figure}

\begin{figure}
\centering
\includegraphics[width=\columnwidth]{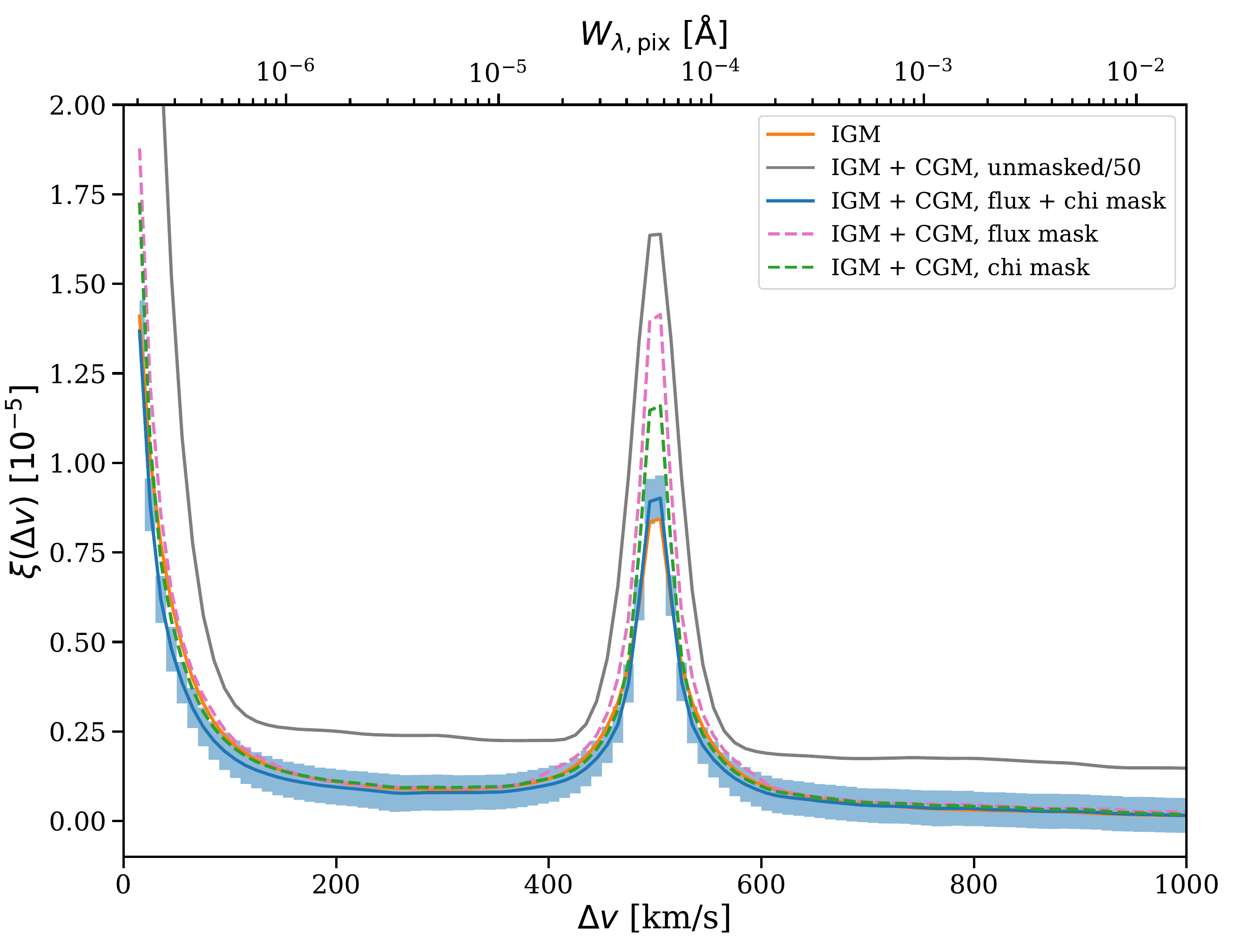}
\caption{Same as Figure \ref{cf_mask_snr20}, but with cutoff thresholds indicated by the vertical magenta lines in Figure \ref{cgm-masking-snr20}. The more aggressive masking is better able to recover the true IGM signal.}
\label{cf_mask_snr20_2}
\end{figure}

In \S \ref{maskingcgm}, we assess the efficiency of masking CGM absorbers on skewers with a random noise of SNR/pix = 50. Here we repeat the masking procedures on lower SNR data, one with SNR/pix = 20. Figure \ref{cgm-masking-snr20} shows the resultant $1-F$ and $\chi$ PDFs, where the cutoff thresholds represented by dashed black lines are placed at where each PDF starts plateauing, $1 - F > 0.15$ and $\chi > 4$. The correlation function of the masked skewers is compared against the true IGM correlation function in Figure \ref{cf_mask_snr20}, where it is apparent that the correlation function of the masked skewers is significantly biased. Figure \ref{cf_mask_snr20_2} shows that by using more aggressive masking thresholds that are placed at the magenta vertical lines in Figure \ref{cgm-masking-snr20} (where $1 - F > 0.06$ and $\chi > 3$), one is able to recover the true IGM correlation function. The more aggressive flux + chi cut removes 38\% of all pixels, compared to 29\% from using the less aggressive cut. Although masking out CGM absorbers is more challenging with lower SNR data by requiring more aggressive masking (otherwise the results will be biased), we can include the values of the cuts in forward modeling and account for them in the inferencing. 

\end{document}